\documentclass[preprint,12pt,number]{elsarticle}

\usepackage{setspace}
\usepackage{amssymb} 
\usepackage{amsmath}
\usepackage{graphicx}
\usepackage{subcaption}
\usepackage{float}
\usepackage[a4paper, total={6in, 9in}]{geometry}

\journal{International Journal of Heat and Mass Transfer}

\begin{document}

\begin{frontmatter}

\title{}
% \title{Improved Approach for the Thermal Characterization of Thin\ Fibers} %% Article title
% \title{Iterative Multi-Parameter Fitting Approach for the Thermal Characterization of Thin\ Fibers} %% Article title

% \title{A novel approach for a high throughput thermal characterization of thin fibers}

\title{A Novel Iterative-Fitting Approach for the Thermal Characterization of Thin Fibers}

\author[label1]{William Spirnock\corref{cor1}} %% Author name
\author[label1]{Alison Hake} %%
\author[label1]{Dihui Wang} %%
\author[label1]{Timothy Sullivan} %%
\author[label1]{Dana He} %%
\author[label1]{Heng Ban} %% Author name

\cortext[cor1]{Corresponding Author Email: heng.ban@pitt.edu}

\affiliation[label1]{organization={Department of Mechanical Engineering and Materials Science, University of Pittsburgh},%Department and Organization
            addressline={3700 O'Hara St}, 
            city={Pittsburgh},
            postcode={15213}, 
            state={PA},
            country={USA}}

%% Abstract
\begin{abstract}

\noindent{The transient electrothermal technique (TET) is widely valued for its simple implementation and versatility in measuring the thermal properties of thin fibers and films. Its accuracy, however, degrades sharply when environmental radiation losses are non-negligible and the surface emissivity is unknown. These conditions are difficult to avoid at the micro-scale, and conventional TET analyses either neglect them or address them through separate measurements that are prone to sequential errors. In this work, a novel implementation of the TET has been developed that determines multiple thermal properties together with the effective surface emissivity from only two transient measurements. An analytical model incorporating thermal radiation and variable heat generation drives a sensitivity-guided, multi-parameter fitting routine, which extracts thermal conductivity and thermal diffusivity from one form of the temperature solution and thermal conductivity and volumetric heat capacity from another. An iterative curve-fitting procedure then recovers the intrinsic properties free of environmental error while requiring a reduced sampling size, significantly minimizing the time to obtain accurate thermal property information. As a verification, an amorphous silicon dioxide (SiO2) optical fiber and a Hi-Nicalon Type S silicon carbide (SiC) fiber were measured, with results agreeing with literature values to within an average error on the order of 10\% across all tested samples. An extensive uncertainty analysis and a numerical uncertainty study were performed to quantify measurement variability and link it to sample geometry selection. This work establishes a TET scheme that, for the first time, simultaneously resolves multiple intrinsic thermal properties, efficiently corrects for environmental radiation errors, and provides quantitative guidance for sample geometry selection.}

\end{abstract}


%\item A novel technique is developed with two separate experiments 
%\item Simulations suggested viability in reducing uncertainty and improving convergence
%\item Physical measurements found limitations if the experiments had inconsistent results
%\item Experiments with consistent results did find a significant reduction in uncertainty
%\item Convergence was improved and achieved for every experiment

%% Keywords
\begin{keyword}
\noindent{Thermal Conductivity, Thermal diffusivity,
Volumetric Heat Capacity, Sensitivity Analysis, Microscale, Electrothermal Technique}


\end{keyword}

\end{frontmatter}



\section{Introduction}

{The improved mechanical and thermal performance of advanced micro- to nanoscale fibers such as carbon fibers/nanotubes \cite{moon_note_2012}, polyacrylonitrile wires \cite{guo_transient_2007}, ultra high molecular weight polyethylene fibers (UHMWPE) \cite{liu_thermal_2015,zhu_novel_2017}, etc play crucial roles in existing and future applications regarding strength and thermal management. However, various difficulties arise when measuring the thermal properties due to difficult sample preparation and environmental sources of error. Additionally, in the case of thin fibers, certain parameters such as volumetric heat capacity can be difficult to determine accurately. This requires the use of conventional methods like differential scanning calorimetry (DSC) which can be time consuming and is very sensitive to sample preparation for complex fiber specimens. To address these limitations, many thermal property measurement methods have been developed specifically for thin fiber geometries. Zhang et al. attached a carbon fiber between a heat sink and hot wire and determined the thermal conductivity by comparing the experimental data to a thermal model \cite{zhang_measurements_nodate}. Xing et al. extended the 3$\omega$ method to a thin fiber \cite{xing_thermal_2014,xing_thermal_2017}. Moon et al. suspended metallic films and microfibers between two electrodes where the steady state electrical resistance is related to the thermal conductivity \cite{moon_note_2012,volklein_method_1984,lee_ultrahigh_2022}. Similarly, Guo \cite{guo_thermal_2007} and Xing \cite{xing_thermal_2014_3omega} developed multiple versions of TET to measure the thermal conductivity and diffusivity of various materials. While some of the shortcomings surrounding thin fiber characterization have been addressed through these advancements, most of these methods are still limited in the properties that can be determined. The methods that do have more thermal property information information available tend to be time consuming and prone to experimental failures.

% Paragraph 2: general overview of what TET is and how thermal properties are determined

{TET is a powerful method with many advantages over other such techniques, having been demonstrated to accurately determine the axial thermal conductivity and diffusivity of both electrically conductive and non conductive thin fibers \cite{guo_thermal_2007,xing_thermophysical_2014}. Compared to other methods like $3\omega$, the measurement time is very short with a maximum of several minutes for a single test depending on the sample thermal properties and geometry \cite{munro_thermal_2016}. This method operates on the principle of Joule heating, in which a sample (with a metallic coating in the case of an electrically non conductive material) is supplied with an electrical current sufficiently large to heat the thin fiber suspended between two electrodes. The average temperature change is measured as a voltage drop across the sample and through nonlinear least squares curve fitting, a thermal model can be used to extract the thermal properties.}

% Paragraph 3: disussion of the SoA for TET...reduced model and current status of Full Model...discuss requirement of large sampling sizes to determine the radiation effect of material with unknown emissivity

{In previous work regarding TET, two formulations are applied, which have been termed the reduced model \cite{guo_thermal_2007} and the full model \cite{xing_analysis_2013}. The reduced model utilizes the one-dimensional transient heat conduction equation with volumetric heat generation and assumes the effects of thermal radiation and convection are negligible. This resulted in high accuracy when measuring samples with either a low surface emissivity such as platinum \cite{xing_thermophysical_2014}, or samples with a small length. However, testing a very short fiber leads to several complications. Experimental difficulties arise from the sample prep process when affixing the sample to the electrodes using silver paste. Encroachment of paste over the fiber reduces fitting accuracy with a more noticeable effect as fiber length decreases \cite{xing_thermal_2014}. Additionally, a smaller fiber length also reduces electrical and thermal resistance. A sufficiently small resistance allows thermal and electrical contact resistance and the paste's inherent resistance to dominate. Follow-up work detailed a revision to the reduced model which incorporated radial convective and radiation heat losses \cite{xing_analysis_2013}. This demonstrated that the error in the measured thermal properties due to thermal radiation is proportional to the square of the length divided by the diameter \cite{munro_thermal_2016}. This error correction had three different implementations, one of which involved a simple additive term representing the thermal radiation effect \cite{xing_thermophysical_2014,xing_analysis_2013}. However, this can only be used in the case of a known surface emissivity. An alternative is the measurement and fitting of multiple samples having different lengths and employment of a linear regression to determine the intrinsic thermal diffusivity and conductivity \cite{liu_characterization_2014,liu_thermally_2014,xu_energy_2014}. A third uses a temperature solution derived from the governing equation which incorporates radiation. In this version, various samples of different lengths (i.e. different thermal radiation magnitudes) are tested and used to fit for either the thermal diffusivity and emissivity or conductivity and emissivity simultaneously. Although this version more effectively accounts for radiation and nonconstant heating outside of an approximated linear regime, the uncertainty is larger than that of the reduced model \cite{munro_thermal_2016} and the thermal diffusivity is reliant on the fit value of conductivity \cite{xing_thermal_2017}. This results in sequential fitting inaccuracies which further propagates error and uncertainty. Additionally, each of the full model implementations requires a large sampling size to confidently determine the intrinsic thermal properties. For non-conductive fibers, a large sample size is compounded with the weeks-long annealing process, which often results in further coating and annealing trials.}

% Paragraph 4: discuss multi-parameter fitting in various methods (3omega, SDTR, PTR, FDTR) and mention the Bo Feng paper for TET and its limitations

{Multi-parameter fitting has been widely incorporated into various thermal characterization methods for both bulk and thin fiber geometries. These include frequency domain thermoreflectance (FDTR) \cite{feser_pump-probe_2014,jiang_new_2022,tang_anisotropic_2021,rodin_simultaneous_2017,schmidt_frequency-domain_2009}, infrared (IR) thermometry \cite{wang_three-dimensional_2025,wang_spatially_2024,sheikh_comparison_2024,hamaoui_spatially_2020,corona_recent_2023,sheikh_photothermal_2025}, and the 3$\omega$ method \cite{xing_thermal_2014,xing_thermal_2017}. Fitting several parameters at once is advantageous in that it avoids propagation of uncertainties by analyzing how the sensitivities of each parameter relate and determine the range of data points best suited for low uncertainty thermal property evaluation. This procedure is more common in the case of bulk materials and has been successfully implemented by Xing et al. in the case of a thin fiber using the same setup as in TET \cite{xing_thermal_2014}. Although Feng et al. demonstrated the potential for multi-parameter fitting with TET \cite{feng_simultaneous_2009} for measuring specific heat and conductivity, this work preceded development of the full model and so had various limitations. Rather than using an exact temperature solution, Feng et al. relied on a numerical model which is largely dependent on the differencing method used and is more susceptible to roundoff errors. Lastly, the authors also did not consider the effects of radiation in the model and did not firmly establish parameter identifiability.}

{In this work, a novel temperature solution of the TET full model is derived which relates the change in voltage with respect to time for two different sets of thermal properties. A sensitivity analysis was employed to explore the potential for two multi-parameter fitting schemes that can eliminate the propagation of error present in former work. This analysis uncovered a tradeoff between two properties which led to the implementation of a novel methodology for extracting the thermal radiation effect when the surface emissivity is unknown. The new methodology is demonstrated on two electrically non-conductive standard samples. All obtained results are in good agreement with existing literature values. Additionally, an uncertainty analysis is performed which reports an improvement from previous full model implementations. An uncertainty numerical study is also performed to guide future work regarding proper geometry selection for minimizing uncertainty of the fit thermal properties.}

% For the first time, this method was demonstrated to be able to independently determine volumetric heat capacity, an important parameter in thermal storage applications.

% This offers a robust alternative to measuring this parameter instead of using conventional methods for measuring specific heat which are more prone to er

% This work also demonstrates the ability to determine volumetric heat capacity, which is an important parameter

%  A new parameter, volumetric heat capacity is determined via TET for the first time. This

\section{Experimental Section}

\subsection{Sample materials and preparation}
{The working principle of TET with regard to sample prep and data collection is well established \cite{liu_characterization_2014,guo_thermal_2007,xing_thermal_2014,xing_thermophysical_2014} and will be briefly described in this and the following section. Two materials are measured in this work to provide validation for the improved methodology: pure silica optical fiber with a diameter ranging from 122-130$\mu$m and HI Nicalon Type S SiC fiber with a diameter range of 10-15$\mu$m. These were selected because they demonstrate the range of TET as spanning two orders of magnitude. As both materials are electrically insulating, several experimental steps were required prior to testing. The pure silica core single-mode optical fibers were obtained from Thorlabs and had an external coating of acrylate which was burned off. The fibers where then cleaned using isopropyl alcohol. The SiC fiber did not undergo the same cleaning to avoid potential surface alteration which could change the measured thermal properties. The fibers were cut to length and suspended between two copper electrodes and fixed at each end with an electrically conductive silver epoxy. A thin layer of gold (10-40 nm) is deposited via sputter coating on the various samples to induce an electrical resistance across the suspended fibers. The post-deposition surface quality and fiber diameter is determined via a Zeiss Sigma500 VP scanning electron microscope (SEM) and the fiber length is measured using a Keyence VHX 7000N microscope. Since the thermal model is sensitive to electrical resistance, the coated samples were placed inside of a furnace and annealed at 100$^\circ$C for seven days to ensure resistance stabilization prior to testing \cite{xing_thermophysical_2014}.}

% add some references in here at various locations

% The general principle of TET will be described in part in the introduction

% should i have an image of the sample being measured?

\subsection{Experimental setup}

% I think i should rework this to instead have the first paragraph listing what everything is for and that it is automated and then the second paragraph can detail the calibration process, the data collection process, and selecting a sufficient current to avoid nonconstant heating...

{A multi-component system was assembled to perform TET measurements and is shown in Figure~\ref{fig:TET_exp_setup}. A Janis ST-100 cryostat is used to house the fiber sample in a vacuum controlled environment. An Agilent Varian TPS-Compact Dry 74FS Turbomolecular Pump is responsible for evacuating the chamber. A LakeShore 335 Temperature Controller is used to moderate the temperature within the chamber. The current across the sample is set by a Keithley DC and AC Current Source and the voltage is measured using a Keithely 3706 Multimeter and Data Acquisition System. Communication of these components is through an instrument control code written in MATLAB. This code performs measurements for the resistance-temperature calibration and for the transient voltage measurements.}

% I feel like i should alter this figure in some way to include what a transient voltage rise looks like

\begin{figure}[H]
    \centering
    \includegraphics[width=1.0\linewidth]{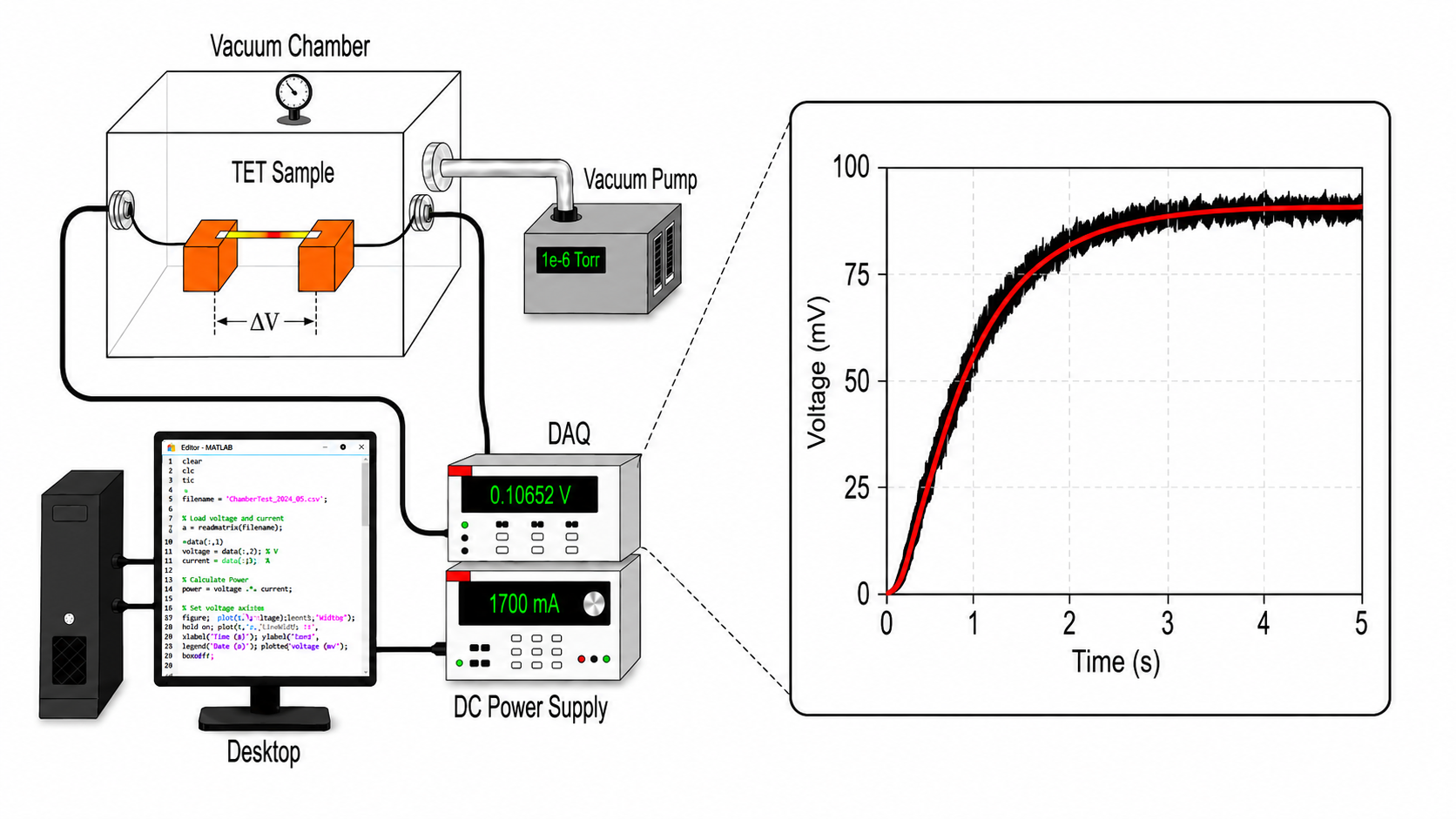}
    \caption{Schematic diagram of the TET configuration.}
    \label{fig:TET_exp_setup}
\end{figure}

% need to mention the calibration somehwere in here

{In this system, the fiber sample is placed inside of a cryostat and is pumped down to a pressure at which convection is negligible ($\leq \, 0.001 \, Pa$) \cite{xing_thermal_2014}. To determine desired properties using the thermal model, a calibration must first be performed to relate resistance/voltage change to a change in temperature. For this calibration, a K-type thermocouple is fixed onto the surface of the copper block close to the fiber sample using the silver paste. The LakeShore 335 Temperature Controller holds the temperature at various set points (eight in total) at which the resistance and temperature are recorded simultaneously by the Keithley 3706 Multimeter. Note that this experiment requires a non-heating current to be used for the resistance measurement. The temperature coefficient of resistance ($\eta$ in this work) was calculated from the slope of the resistance-temperature (R-T) curve \cite{munro_thermal_2016}. The R-T plots can be seen in the supplementary materials document. This process must be performed for each sample, regardless if the material type is unchanged, to account for defects and variations in coating thickness. Once the calibration is complete, a step current is applied across the sample which results in a temperature rise due to joule heating. This response is measured as a transient voltage drop across the sample. This current must be carefully selected to avoid significant contributions from nonconstant heating \cite{xing_thermophysical_2014}. Small heat losses are approximated by the thermal model but nonlinearities arise at high temperatures. It was found that $\Delta T_{fiber}=5K$ resulted in an acceptable signal to noise ratio and did not have noticeable variable heating losses. The temperature was determined from the previously mentioned calibration process.}

% {Note that various parameters are determined prior to data collection. Previous work [citation here] has demonstrated that a substantial temperature rise of the sample can lead to nonconstant heating losses. Even though this is factored into the thermal model, too significant of losses can diverge from what the model predicts. There are two primary options to account for this. The first is to collect data at various currents and fit for the thermal properties. If the thermal properties remain constant, nonconstant heating is negligible.}

\subsection{Theory}

% Existing work simplifies this derivation via two primary forms to account for potential heat losses due to environmental sources of error such as thermal radiation. Xing et al non-dimensionalizes the governing equation and separates the solutions of thermal conductivity and thermal diffusivity in which thermal diffusivity is determined via normalized transient voltage rise and conductivity from the change in resistance \cite{xing_analysis_2013}. Other work simplifies the governing equation and incorporates thermal radiation as an additive correction to the measured value of either property \cite{lin_significantly_2013}. These methods have proven to be effective, however, both require a large sampling size for accurate error estimation and exhibit an interdependence of the fit parameters which can further propagate errors.

Figure~\ref{fig:TET_sample} shows a close-up of the TET sample. The fiber is suspended between two electrodes and attached to each end by thermal paste to reduce thermal/electrical contact resistance between the sample and heat sink. The system can be modeled as a one-dimensional heat equation with volumetric heat generation assuming $\frac{L}{D}>>10$, making any radial temperature gradients negligible \cite{xing_analysis_2013}.

\begin{figure}[H]
    \centering
    \includegraphics[width=0.9\linewidth]{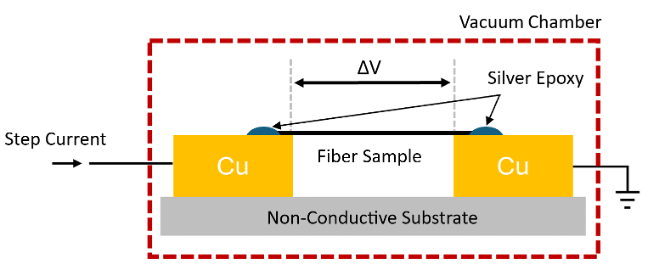}
    \caption{Illustration of the TET sample placed inside of the vacuum chamber for testing.}
    \label{fig:TET_sample}
\end{figure}

{However, to reduce propagation of errors due to sequential parameter fitting and to expand upon the properties that can be determined using TET, the governing equation in \cite{xing_analysis_2013} is used to derive an alternate temperature solution. The transient one-dimensional heat conduction equation for the TET method is written as:}

\begin{equation}
\frac{1}{\alpha}\frac{\partial T(x,t)}{\partial t}
=
\frac{\partial^2 T(x,t)}{\partial x^2}
-
H_e^2\,[T(x,t)-T_0]
+
\frac{q'''}{k}
\label{eq:TET_gov_eqn}
\end{equation}

{In Equation~\eqref{eq:TET_gov_eqn}, $q^{'''}=\frac{IV_{0}}{V_{s}}$ represents the volumetric heat generation term where $I$ is the current, $V_{0}$ is the initial voltage, and $V_s = \frac{\pi D^2 L}{4}$ is the volume of the fiber, where $L$ and $D$ are the fiber length and diameter, respectively. The thermal diffusivity is represented by $\alpha$ and the thermal conductivity by $k$. The term $H_{e}^{2}=\frac{(h_c+h_r)A_l-\eta V_0I}{kV_s}$ is the effective heat transfer coefficient which accounts for heat losses from the surface of the fiber and from temperature dependent resistance change of the sample and coating. Since the experiment takes place in vacuum, convection is negligible ($h_c=0$). Assuming a gray body, and small change in fiber temperature during the experiment, the radiation heat transfer is linearized using a Taylor expansion to obtain $h_r \approx 4\epsilon\sigma T_{0}^{3}$. The emissivity is represented by $\epsilon$, $T_0$ is the sample initial temperature, and  $\sigma=5.67*10^{-8} W\,m^{-2}\,K^{-4}$ is the Stefan-Boltzman constant.}

{Assuming constant temperature boundary conditions due to the heat sinks, the temperature solution is determined by applying Green's function general solution approach for a one-dimensional nonhomogeneous heat equation \cite{book} to arrive at the spatially dependent and transient full model TET solution which involves heat losses via radiation and nonconstant heating.}

\begin{equation}
\theta(x,t)=\frac{4\alpha q'''}{Lk}
\sum_{m=1}^{\infty}
\frac{1-e^{-\alpha\left(\lambda_m^2+H_e^2\right)t}}
{\alpha\left(\lambda_m^2+H_e^2\right)\lambda_m}
\sin(\lambda_m x)
\label{eq:TET_temp_soln}
\end{equation}

{where $\lambda_{m}^{2}=\frac{(2m-1)^2\pi^2}{L^2}$ and $\theta(x,t)=T(x,t)-T_0$. However, since the signal is read from the heat sinks, the system response is the spatially averaged temperature across the fiber. This temperature is measured as a voltage drop across the sample. The temperature can be converted to voltage through established resistance-temperature correlations to arrive at the alternate solution for the TET full model.}

\begin{equation}
\Delta V(t)
=
\frac{8\eta I V_0^{2} L^{2}}{kV_s}
\sum_{m=1}^{\infty}
\frac{
1-\exp\!\left(-\frac{\left((2m-1)^{2}\pi^{2}-L^{2}H_{e}^{2}\right)\Gamma_{1,2} t}{L^{2}}\right)
}{
(2m-1)^{4}\pi^{4}-L^{2}H_{e}^{2}(2m-1)^{2}\pi^{2}
}
\label{eq:TET_volt}
\end{equation}

% {Equation~\eqref{eq:TET_volt} differs from previous TET derivations \cite{xing_analysis_2013}, where thermal conductivity is first estimated from the steady-state resistance and thermal diffusivity is then fit from the normalized transient voltage. Because diffusivity depends on conductivity, uncertainty in the conductivity measurement can propagate into the diffusivity result. In the present formulation, conductivity is fit simultaneously with either volumetric heat capacity or thermal diffusivity from the same voltage-response equation. Although the thermal-property interdependence is not eliminated, the multi-parameter fit allows the coupling between fitted parameters to be quantified through their covariance. In 
% Equation~\eqref{eq:TET_volt}, the thermal conductivity appears in both the transient and steady state terms. In the transient term, $\Gamma_{1,2}$ represents two distinct formulations, one in which thermal conductivity and volumetric heat capacity appear, and another in which only thermal diffusivity is present. More specifically, we define these as $\Gamma_1=k/(\rho C_p)$, and $\Gamma_2=\alpha$. Although $\alpha=k/(\rho C_p)$, the presence of thermal conductivity throughout the derived voltage-response signifies a more complex relationship between the three parameters. This work provides a detailed comparison of both implementations with regard to their accuracy and uncertainty to assess this relationship and provide guidance as to which combination of parameters is most suitable for thin fiber characterization.
% }

Equation~\eqref{eq:TET_volt} differs from previous TET approaches \cite{xing_analysis_2013}, in which thermal conductivity is first estimated from the steady-state resistance and thermal diffusivity is subsequently obtained by fitting the normalized transient voltage. Because the diffusivity fit depends on the previously determined conductivity, uncertainty in the conductivity measurement may propagate into the resulting diffusivity. In the present formulation, thermal conductivity is instead fit simultaneously with either volumetric heat capacity or thermal diffusivity using the complete voltage response, allowing the covariance and uncertainty associated with the fitted parameters to be evaluated directly.

{Two alternative formulations of the transient term are therefore considered. In the first, the transient decay parameter is written as $\Gamma_1=k/(\rho C_p)$, such that thermal conductivity and volumetric heat capacity are treated as independent fitting parameters. In the second, it is written as $\Gamma_2=\alpha$, and thermal conductivity and thermal diffusivity are fit directly. Although $\alpha=k/(\rho C_p)$, these two fitting schemes are not necessarily equivalent from a parameter-estimation perspective. Thermal conductivity does not appear only through the diffusivity relation; it also enters independently at multiple locations within both the transient and steady-state components of Equation~\eqref{eq:TET_volt}. Consequently, substituting $k/(\rho C_p)$ for $\alpha$ changes the sensitivities, correlations, and uncertainty structure of the fitted parameters. This work compares the accuracy and uncertainty of both implementations to clarify these parameter relationships and determine whether fitting $(k,\rho C_p)$ or $(k,\alpha)$ is more suitable for the thermal characterization of thin fibers.} Throughout this document, the two formulations for fitting $(k,\rho C_p)$ and $(k,\alpha)$ will be referred to as the $\Gamma_1$ and $\Gamma_2$ fitting schemes, respectively.

\section{Results and discussion}

\subsection{Parameter identifiability}

A sensitivity analysis was performed to evaluate the influence of each fitting parameter on the transient voltage response and to assess the potential for multi-parameter fitting. The sensitivity is defined as \cite{tang_anisotropic_2021}

\begin{equation}
S_{x_u} = \frac{\partial \ln(\Delta V)}{\partial \ln(x_u)}
\end{equation}

\noindent{}where $\Delta V$ is the measured transient voltage response and $x_u$ is the perturbed parameter. In this work, the parameters of interest are $k$, $\rho C_p$, $\alpha$, and $\epsilon$. The inclusion of $\rho C_p$ follows from the $\Gamma_1=k/(\rho C_p)$ fitting scheme, while $\alpha$ is considered in the $\Gamma_2=\alpha$ scheme. Other quantities, including $L$, $I$, $V_0$, $D$, $\eta$, and $T_0$, also affect the voltage response; however, these are treated as measured inputs rather than fitted parameters.

Previous TET studies have shown that radiation losses can be represented by the dimensionless parameter $\xi = \frac{16 \epsilon \sigma T_0^3 L^2}{kD}$, which increases with sample length and affects the accuracy of reduced-model fitting \cite{xing_analysis_2013,munro_thermal_2016}. Therefore, the effect of sample length on parameter sensitivity was examined using our novel full-model formulation.

Sensitivity calculations were first performed for a standard $SiO_2$ fiber using two sample lengths, $5\,\mathrm{mm}$ and $25\,\mathrm{mm}$, representing the lower and upper practical limits of the experimental system. The manufacturer-reported diameter of $124.5\,\mu\mathrm{m}$ was used, along with an effective emissivity of 0.5 and a calibrated temperature coefficient of resistance of $1.38\times10^{-3}\,\mathrm{K}^{-1}$. The current was selected to be near zero to eliminate variable heating effects to isolate the impact of thermal radiation while maintaining an adequate signal-to-noise ratio.

\begin{figure}[H]
\centering

\begin{subfigure}[t]{0.33\textwidth}
\centering
\includegraphics[width=\linewidth]{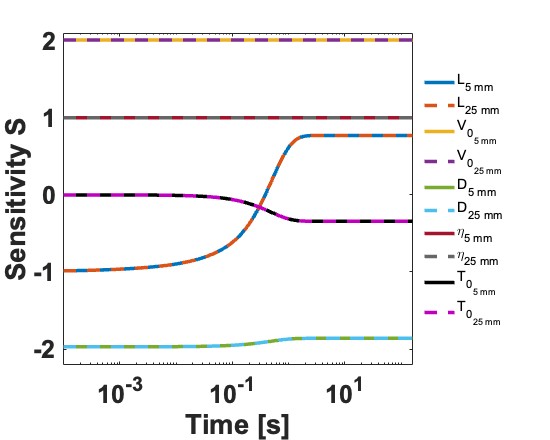}
\subcaption{}\label{fig:measured inputs}
\end{subfigure}
\begin{subfigure}[t]{0.33\textwidth}
\centering
\includegraphics[width=\linewidth]{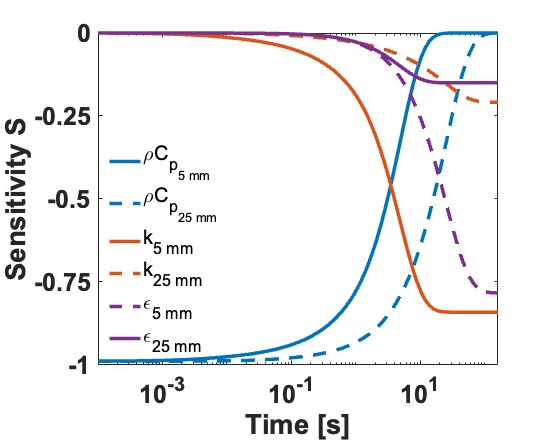}
\subcaption{}\label{fig:pCp_5mm}
\end{subfigure}\hfill
\begin{subfigure}[t]{0.33\textwidth}
\centering
\includegraphics[width=\linewidth]{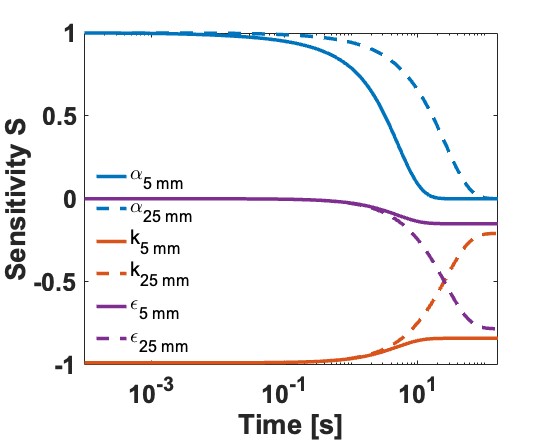}
\subcaption{}\label{fig:pCp_25mm}
\end{subfigure}

\caption{Line plots for the $SiO_2$ (a) sensitivity to the measured inputs, (b) sensitivities for the $\Gamma_1$ scheme, and (c) sensitivities for the $\Gamma_2$ scheme.}
\label{fig:2D sens plots silica}
\end{figure}

The sensitivity of the measured inputs were found to have no dependence on the radiation effect as shown in Figure~\ref{fig:measured inputs}. That is, the trends and magnitudes remain unchanged from the shorter and longer sample. We also see that other than length, these parameters are nearly constant over the measurement time, meaning they have a fixed impact at any given point in time. The sensitivities for the thermal properties are also plotted for both sample lengths with their trends showing that the fitted parameters have distinct effects on the transient response, indicating that simultaneous fitting is potentially feasible. Unlike the measured inputs, there is more variation in the parameter sensitivities at different radiation magnitudes for both fitting schemes. 
The sensitivities of $\rho C_p$ and $\alpha$ are merely shifted with increased $\xi$, while the sensitivity to $k$ moves toward zero in both Figure~\ref{fig:pCp_5mm} and Figure~\ref{fig:pCp_25mm}. The sensitivities to $k$ and $\epsilon$ also exhibit a tradeoff between the low and high radiation samples with regard to their magnitudes. Furthermore, for small $\xi$, the $\Gamma_1$ scheme shows greater variation in sensitivity to $k$ for the change in time as opposed to $\Gamma_2$ which is relatively fixed over time as in the case of the measured inputs. At early times the sensitivities vary in magnitude from $S_{\rho C_p} \approx -1$ and $S_k \approx 0$. In this range, the transient response is highly sensitive to $\rho Cp$ and less sensitive to $k$. However, at later times, their values approach $S_k \approx -0.84$ and $S_{\rho C_p} \approx 0$, meaning there is a reduced sensitivity to $\rho Cp$ and a dominant effect from $k$. The sensitivity overlap in the intermediate time range demonstrates the complex relationship of these two parameters which extends beyond the simple relationship defined by thermal diffusivity. For $\Gamma_2$, we see a slightly different relationship between $\alpha$ and $k$. At early times, the sensitivities vary from $S_{\alpha} \approx 1$ and $S_k \approx -1$. They share nearly the same sensitivity magnitudes for the initial transient until they begin to approach $S_{\alpha} \approx 0$ and $S_k \approx -0.84$. Unlike in $\Gamma_1$, the sensitivity to conductivity does not reduce to zero and shows less variation in $S_k$ overall. However, for the majority of the transient region, there is a clear separation in the magnitude of $S_\alpha$ and $S_k$, indicating $\Gamma_2$ has high potential for multiparameter fitting as well. However, as $\xi$ increases, $S_{k}$ and $S_{\alpha}$ have a mirrored trend about the time axis. The sensitivity vectors become nearly collinear. This indicates that for the $\Gamma_2$ scheme, the parameter identifiability is increasingly ill-conditioned as $\xi$ becomes larger. Therefore, while $\Gamma_2$ may be capable of multi-parameter fitting for low radiation samples, we would expect uncertainty to increase and accuracy to decrease beyond a certain level. This analysis demonstrates the unique behavior of each scheme and suggests that when the approximate magnitude of $\xi$ is known, both fitting schemes have similar accuracy and uncertainty. Although across all possible $\xi$ values, $\rho C_p$ and $k$ may have less parameter coupling and thus could be the safer option. 

The same analysis was performed for a $13\,\mu\mathrm{m}$ HI-Nicalon Type-S SiC fiber, where the diameter was determined from SEM measurements.

\begin{figure}[H]
\centering

% \begin{subfigure}[t]{0.50\textwidth}
% \centering
% \includegraphics[width=\linewidth]{Images/Sens_Gamma2_others_silica.jpg}
% \subcaption{}\label{fig:pCp_25mm_SiC}
% \end{subfigure}
\begin{subfigure}[t]{0.50\textwidth}
\centering
\includegraphics[width=\linewidth]{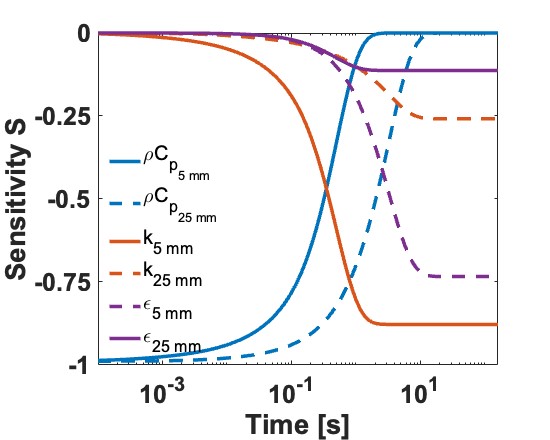}
\subcaption{}\label{fig:pCp_5mm_SiC}
\end{subfigure}\hfill
\begin{subfigure}[t]{0.50\textwidth}
\centering
\includegraphics[width=\linewidth]{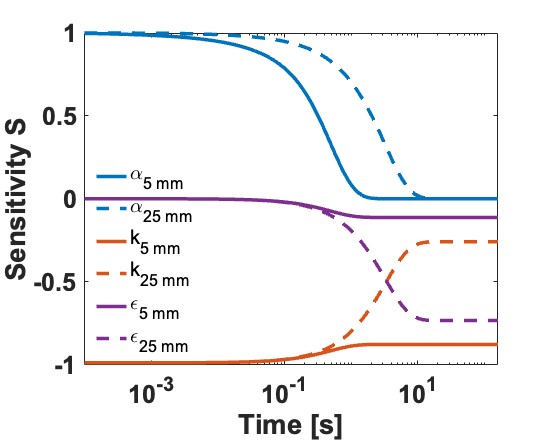}
\subcaption{}\label{fig:pCp_25mm_SiC}
\end{subfigure}

\caption{Line plots for the $SiC$ (a) sensitivity to the measured inputs, (b) sensitivities for the $\Gamma_1$ scheme, and (c) sensitivities for the $\Gamma_2$ scheme.}
\label{fig:2D sens plots SiC}
\end{figure}

The SiC trends and tradeoffs are nearly identical to silica with the single difference being their time-dependent magnitudes resulting from different thermal properties. Both samples and schemes suggest shorter samples provide greater sensitivity to the thermal properties of interest, while longer samples reduce sensitivity to $k$ and increase sensitivity to $\epsilon$ which would increase uncertainty for the $k$ measurement.

% To generalize the length dependence, the sensitivities were also mapped as functions of the radiation parameter $\xi$ and transient heating time for various materials. While there is variation in the trend with respect to time, due to property dictated rise times, the general trend of the sensitivity landscape remains constant. These results can be seen in the supplementary materials document.

The sensitivity plots demonstrate the potential for multi-parameter fitting, however, both schemes show signs of interdependence which can affect fitting accuracy. More specifically, $\Gamma_1$ shows a possible linear dependence of $S_k$ and $S_\epsilon$ while $\Gamma_2$ shows similar trends for $S_\alpha$ and $S_k$. As mentioned, this becomes more pronounced at longer lengths and thus larger $\xi$ values. Similar concerns have been identified in previous work regarding multi-parameter fitting \cite{tang_anisotropic_2021,jiang_new_2022}. Tang et al. and Jiang et al. both demonstrated that upon applying a nonlinear least squares curve-fitting algorithm, the sum of squared errors (SSE) can be plotted with respect to the fitted parameters to visualize the cost landscape. 

{To identify the SSE trends, a non-linear least squares curve-fitting code was developed in MATLAB based on the Levenberg-Marquardt algorithm. The code was used to generate artificial data in the form of a transient voltage rise. The surface emissivity was assumed to be a known value. After fitting was completed, the SSE was plotted with respect to the property space.}

\begin{figure}[H]
\centering

\begin{subfigure}[t]{0.32\textwidth}
  \centering
  \includegraphics[width=\linewidth]{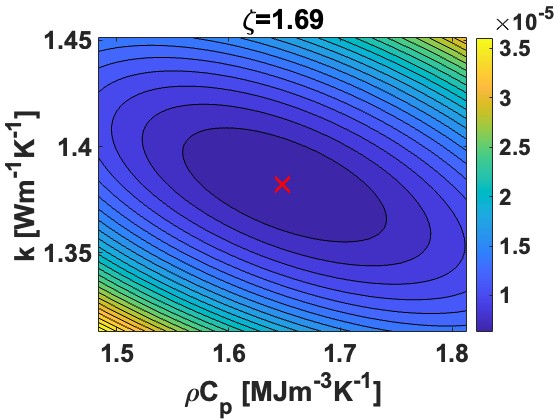}
  \subcaption{}\label{fig:5mm contour}
\end{subfigure}\hfill
\begin{subfigure}[t]{0.32\textwidth}
  \centering
  \includegraphics[width=\linewidth]{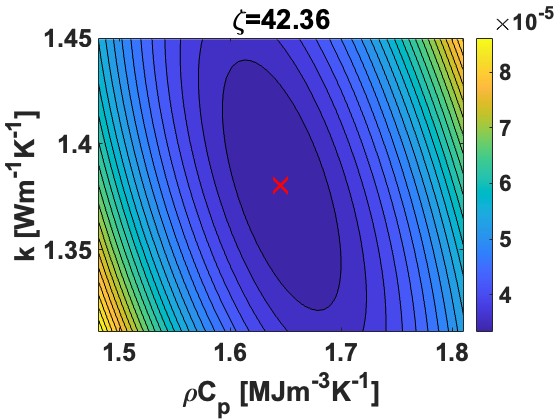}
  \subcaption{}\label{fig:25mm contour}
\end{subfigure}\hfill
\begin{subfigure}[t]{0.32\textwidth}
  \centering
  \includegraphics[width=\linewidth]{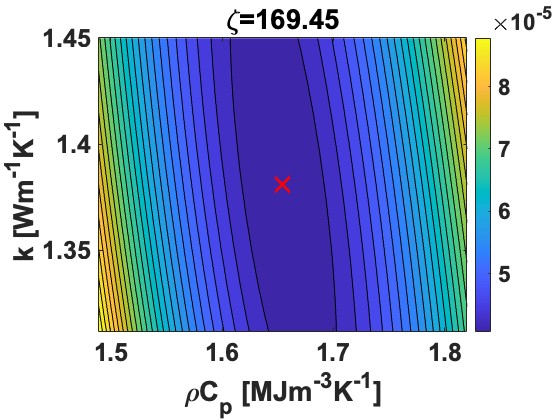}
  \subcaption{}\label{fig:50mm contour}
\end{subfigure}

\caption{$\Gamma_1$ SSE contour plots of $SiO_2$ for (a) 5 mm (b) 25 mm (c) 50 mm}
\label{fig:SSE plots}
\end{figure}

{The generated data correspond to a silica sample (as used for the sensitivity analysis) for three different lengths: 5 mm, 25 mm, and 50 mm. The 50 mm length was included to demonstrate the noticeable change in the cost landscape at a significant thermal radiation effect. Figure~\ref{fig:SSE plots} demonstrates that a small $\xi$ results in a convex shape and thus is well-conditioned for multi-parameter fitting. However, as $\xi$ increases, the cost landscape becomes increasingly parabolic with an elongated valley, meaning the thermal properties cannot be untangled without significantly risking accuracy and uncertainty. These results demonstrate that the thermal properties will have higher accuracy and lower uncertainty by testing shorter length samples, corroborating the sensitivity analysis. The SSE results for $\Gamma_2$ were similar, with slight variations; the corresponding plots can be found in the supplementary materials document.

These trends also demonstrate why simultaneous fitting is preferable to a sequential approach. In sequential fitting, one property is first determined and then treated as a fixed input when fitting the second property, which can obscure parameter coupling and transfer error directly from the first fit into the second. In contrast, simultaneous fitting allows the coupled influence of the parameters to appear directly in the SSE landscape and covariance matrix. Therefore, the interdependence is not eliminated, but it can be quantified and used to assess when the fitted parameters can be resolved reliably.

\subsection{Iterative approach for thermal radiation error correction}

{The ability to fit multiple parameters simultaneously has been established for shorter length fiber samples. However, in most cases, the emissivity is a parameter with non-negligible impact that must be determined. Recall from Figure~\ref{fig:2D sens plots silica} and Figure~\ref{fig:2D sens plots SiC} that a tradeoff exists between $S_k$ and $S_{\epsilon}$ for both $\Gamma_1$ and $\Gamma_2$. When $\xi$ is small, the sensitivity to conductivity is much larger than that of emissivity, and this is reversed as $\xi$ increases. This indicates that while a low $\xi$ can be used to fit for the thermal properties with reduced error from an unknown emissivity, a large $\xi$ can be used to determine emissivity with a reduced impact from unknown thermal properties. Similar methodologies have been employed in previous work. Jiang et al. \cite{jiang_new_2022} capitalized on a sensitivity tradeoff using spatial domain thermoreflectance (SDTR) in which two data sets were used to fit for phase and laser spot size in an iterative curve fitting procedure. Our work adopted a similar approach which is used to determine the thermal properties and the emissivity from only two data sets. In this process, we took advantage of our control in $\xi$ by using a short and a long fiber sample to obtain data sets with distinct sensitivities for the three parameters in question. To verify this methodology, a SiC sample was tested for a 4.84 mm and 24.6 mm sample.}

\begin{figure}[H]
    \centering
    \includegraphics[width=1\linewidth]{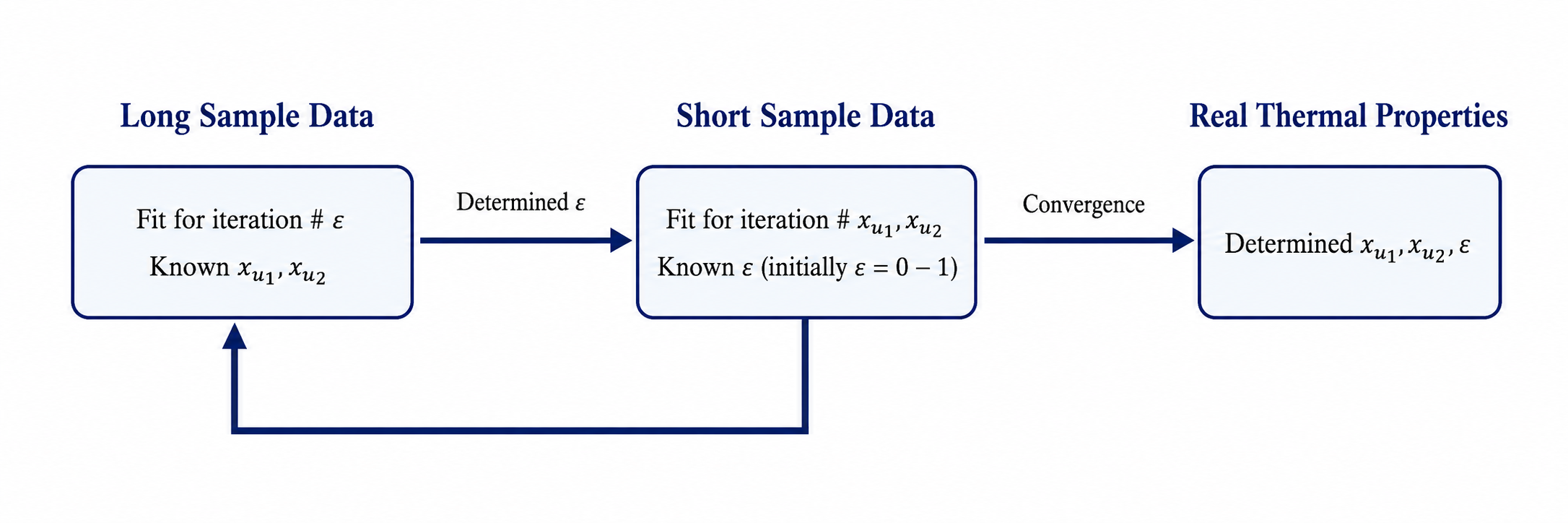}
    \caption{TET iterative curve fitting procedure}
    \label{fig:iter_flow}
\end{figure}

{Figure~\ref{fig:iter_flow} describes the process used to determine the three parameters for both fitting schemes. A transient voltage curve is collected for a short and long sample of the same material. First, guessed values for the thermal properties in $\Gamma_1$ and $\Gamma_2$ are used to fit an initial value for emissivity. This value is then used as a known in the proceeding code which fits for the thermal properties. The initial fit for each of these parameters will be larger or smaller than the real values depending on the initial guess. The fit values for $\rho C_p$ or $\alpha$ and $k$ are then updated as the known values for the emissivity fitting. This process continues iteratively until a user prescribed convergence criteria is met. The converged values represent the thermal properties unaffected by thermal radiation, i.e. the intrinsic thermal properties. To ensure the iterative process is independent of the initial assumptions for $k$, $\rho C_p$, and $\alpha$, their magnitudes were varied using various multiples of the literature values. Each alteration resulted in the same magnitudes, with variations only in convergence time.}

% should i include this in the actual paper?

\begin{figure}[H]
\centering

\begin{subfigure}[t]{0.33\textwidth}
  \centering
  \includegraphics[width=\linewidth]{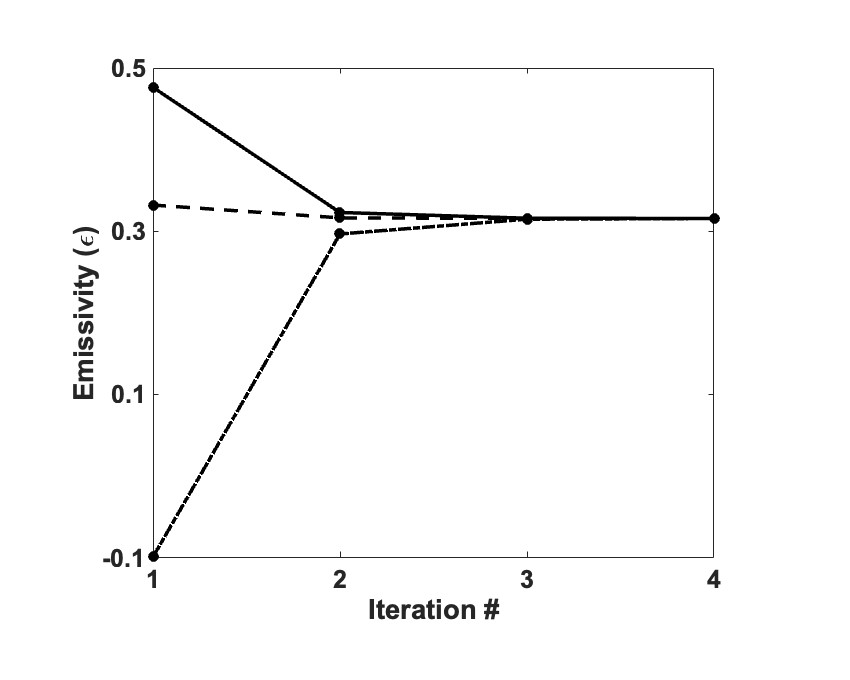}
  \subcaption{}\label{fig:eps_conv}
\end{subfigure}\hfill
\begin{subfigure}[t]{0.33\textwidth}
  \centering
  \includegraphics[width=\linewidth]{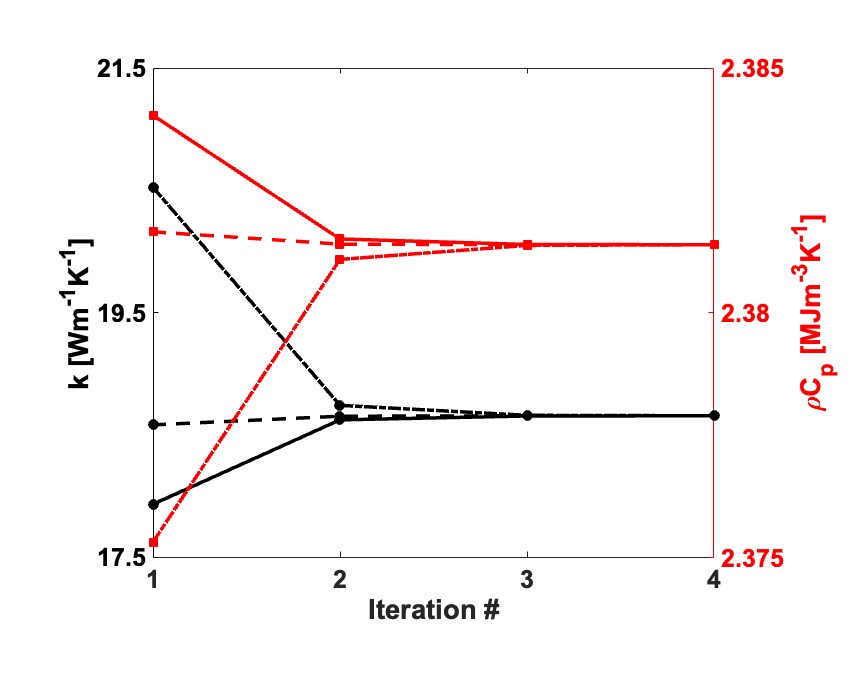}
  \subcaption{}\label{fig:k_pCp_conv}
\end{subfigure}\hfill
\begin{subfigure}[t]{0.33\textwidth}
  \centering
  \includegraphics[width=\linewidth]{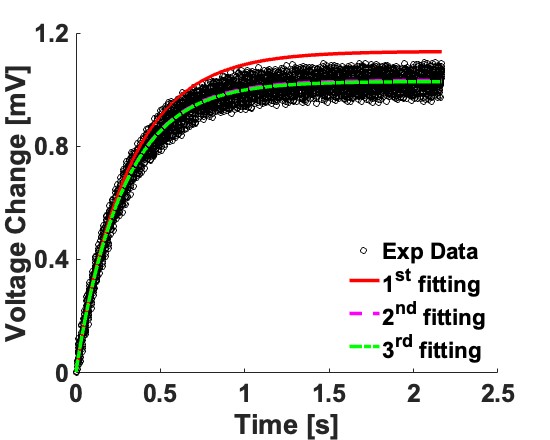}
  \subcaption{}\label{fig:iter_dem}
\end{subfigure}

\caption{Iterative TET convergence plots for (a) $\epsilon$ (b) $k$ and $\rho C_p$ and (c) a plot with artificial data demonstrating how the final fit curve approaches the converged value for each iterative step.}
\label{fig:iter_example}
\end{figure}

{Figure~\ref{fig:iter_example} depicts the convergence of the three properties across four itetations for the $\Gamma_1$ scheme. There is no significant difference in the convergence trend for $\Gamma_2$. Note that the trend plateaus fairly quickly around the third or fourth iteration. Extensive numerical simulations have been performed to determine the accuracy and uncertainty of this approach for various combinations of thermal properties and sample to sample $\xi$ ratios and will be discussed in a later section. 
Also, note the plot shows different initial fit values for the emissivity to again demonstrate the assumed $k$, $\rho Cp$, and $\alpha$ do not impact the final fitting value, but only the time until convergence is achieved. Therefore, a global minimum is being found as opposed to a local minimum. This procedure was used to test various combinations of fit parameters, including those mentioned here. The improved methodology now only requires testing two samples as opposed to ten or more in previous approaches for the TET full model \cite{xing_thermal_2014_3omega,munro_thermal_2016,xing_thermophysical_2014}. This novel methodology, based on a multi-parameter fitting approach, can both reduce the time for characterization and provide more overall information compared to previous implementations.}

\subsection{Experimental validation}

{Experimental testing was performed to validate this improved methodology. Twelve short and three long samples of $SiO_2$ optical fiber were tested with diameters in the range 122-130 $\mu$m. Approximately 20 nm of gold was deposited on each sample. Eight short and three long samples of HI Nicalon Type S SiC fiber were also tested. The SiC fibers had a much smaller diameter in the range of 10-15 $\mu$m and so were coated with 40 nm of gold to ensure strong electrical connection. The gold effect of each sample was calculated and found to be negligible. One short sample from both the $SiO_2$ and SiC sets were removed based on Chauvenet's criteria for outliers. Prior to testing, sample annealing was performed to reduce resistance drift for determination of the temperature coefficient of resistance in which the samples were maintained at 100 $^\circ$C for one week. As discussed in previous sections, a resistance temperature calibration was performed to determine $\eta$ and this value was then used to select a current for each sample corresponding to an approximate 5 K temperature rise. For each sample, 30 trials were run and the reported value is the average. This was done to better capture the variability of the improved method. These trials were also used to determine the standard deviation, which is required for the parameter uncertainty. Each short sample is fit using three different long samples to demonstrate the regressed values have little dependence on fiber to fiber variations of the longer sample.}

\begin{figure}[H]
\centering

\begin{subfigure}[t]{0.50\textwidth}
  \centering
  \includegraphics[width=\linewidth]{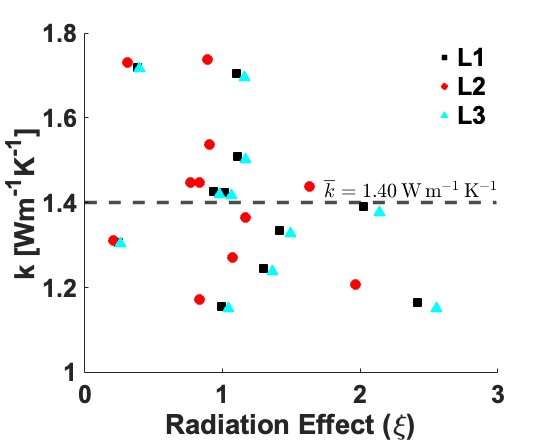}
  \subcaption{}\label{fig:k}
\end{subfigure}\hfill
\begin{subfigure}[t]{0.50\textwidth}
  \centering
  \includegraphics[width=\linewidth]{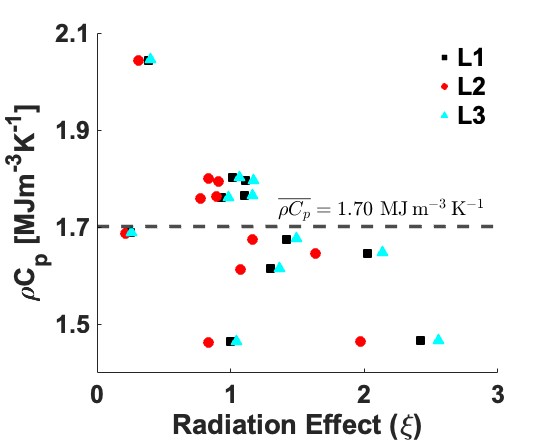}
  \subcaption{}\label{fig:pCp}
\end{subfigure}\hfill
\begin{subfigure}[t]{0.50\textwidth}
  \centering
  \includegraphics[width=\linewidth]{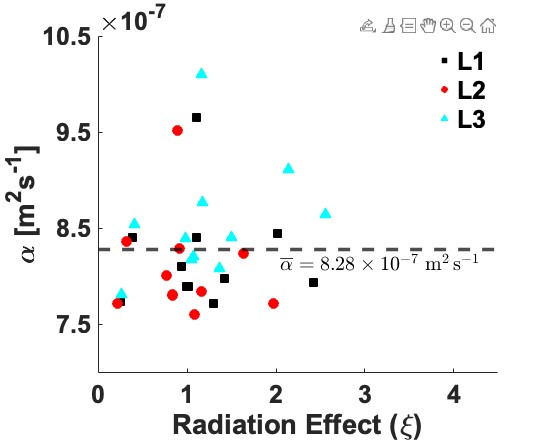}
  \subcaption{}\label{fig:alpha}
\end{subfigure}

\caption{$Si0_{2}$ validation results for (a) $k$ from the $\Gamma_1$ fitting scheme, (b) $\rho C_p$ from the $\Gamma_1$ fitting scheme, (c) $\alpha$ from the $\Gamma_2$ fitting scheme.}
\label{fig:Si02 results}
\end{figure}

{Figure~\ref{fig:Si02 results} shows the results for the $SiO_2$ samples. Note that two samples resulted in approximately 25\% error for thermal conductivity, however, it was kept as it did not fall within the range of removal for outliers. Regardless, the majority of the samples tested fall within 15\% of literature values for $\rho C_p$ and $k$ for the $\Gamma_1$ scheme with an average error of 7.66\% and 10.52\% for $\rho C_p$ and $k$, respectively. The average thermal properties across all samples resulted in an error of 3.11\% and 1.76\% for $\rho C_p$ and $k$, respectively. The literature values are reported as $\rho C_{{p}_{lit}}=1.65*10^{6}\,MJm^{-3}K^{-1}$}, $k_{lit}=1.38\,Wm^{-1}K^{-1}$, and $\alpha_{lit}=8.36*10^{-7}\,m^{2}s^{-1}$. The average uncertainty across the tested samples was $U_{\rho C_{p}}=4.17\%$ and $U_k=4.56\%$. Detailed information regarding uncertainty for each trial can be found in the supplementary materials document. The $\Gamma_2$ scheme returned nearly identical results for thermal conductivity with a higher accuracy for $\alpha$ compared to $\rho C_p$. The majority of the samples fell within 15\% of literature values with an average error of 4.94\% for $\alpha$ and 10.60\% for $k$. The average thermal properties across all samples resulted in an error of 0.91\% and 2.31\% for $\alpha$ and $k$, respectively. The diffusivity had an average uncertainty of $U_{\alpha}=3.19\%$ and the thermal conductivity had $U_{k}=4.61\%$. Again, recall that the fit emissivity is an effective value of the fiber and transducer material and so it cannot be compared to the literature. Rather, it is only used to determine the other parameters in the case of electrically non-conductive materials.}

% There is variation in the accepted value for volumetric heat capacity and thermal diffusivity for fused silica \cite{degallaix_thermal_2007,combis_evaluation_2012}, however, the values used here are taken from previous work regarding TET for consistency and better comparison between methods \cite{xing_thermophysical_2014}. 

\begin{figure}[H]
\centering

\begin{subfigure}[t]{0.50\textwidth}
  \centering
  \includegraphics[width=\linewidth]{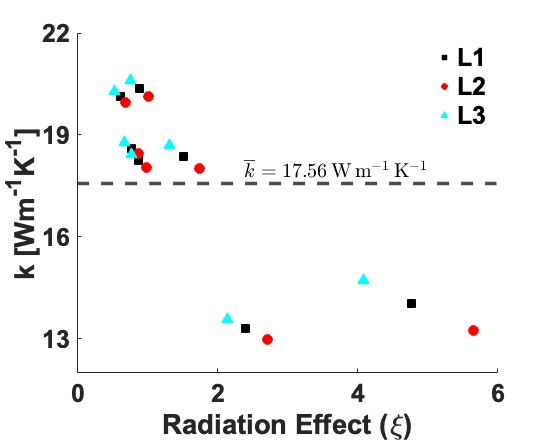}
  \subcaption{}\label{fig:kmeas}
\end{subfigure}\hfill
\begin{subfigure}[t]{0.50\textwidth}
  \centering
  \includegraphics[width=\linewidth]{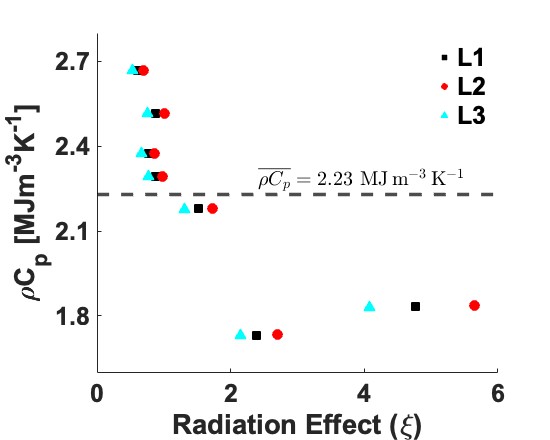}
  \subcaption{}\label{fig:pCpmeas}
\end{subfigure}\hfill
\begin{subfigure}[t]{0.50\textwidth}
  \centering
  \includegraphics[width=\linewidth]{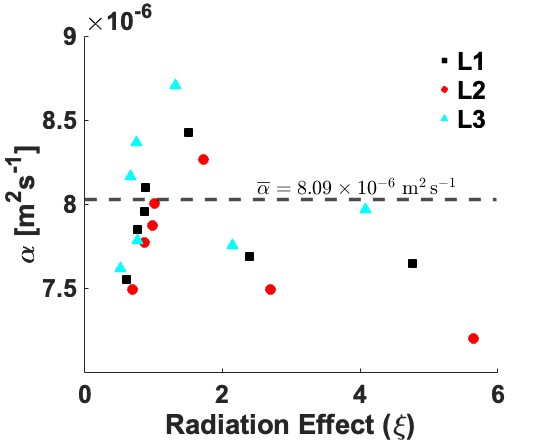}
  \subcaption{}\label{fig:alpha}
\end{subfigure}

\caption{SiC validation results for (a) $k$ from the $\Gamma_1$ fitting scheme, (b) $\rho C_p$ from the $\Gamma_1$ fitting scheme, (d) $\alpha$ from the $\Gamma_2$ fitting scheme.}
\label{fig:SiC results}
\end{figure}

{Figure~\ref{fig:SiC results} shows the results for the SiC fiber samples. The majority of the samples fall within 16\% of the literature values, $\rho C_{{p}_{lit}}=2.17\,MJm^{-3}K^{-1}$, $k_{lit}=18.4\,Wm^{-1}K^{-1}$, $\alpha=8.48*10^{-6}\,m^{2}s^{-1}$ \cite{coi_ceramics_hi-nicalon_type_s}. For the $\Gamma_1$ scheme, the average error was found to be 12.84\% for $\rho C_p$ and 10.59\% for $k$. The error for the average thermal properties across all samples was found to be 2.66\% and 4.55\% for $\rho C_p$ and $k$, respectively. As was the case for the $Si0_2$ samples, $\alpha$ showed an improved accuracy with an average error of 6.93\% for $\alpha$ and 10.08\% for $k$. The error for the average thermal properties of all samples was 6.93\% for $\alpha$ and 4.66\% for $k$. The average uncertainty was for the $\Gamma_1$ scheme found to be $U_{\rho C_{p}}=11.05\%$, $U_k=12.77\%$. For the $\Gamma_2$ scheme, these values are $U_{\alpha}=4.44\%$ and $U_k=12.83\%$. Additional detailed data for each sample is tabulated in the supplementary materials document. Notice that there is a larger uncertainty associated with the SiC fiber when compared to the $SiO_2$ fibers. This is due to the overall higher $\xi$ based on the material thermal properties and fiber diameter. The fiber diameter measurement also contributes to this larger uncertainty as the SiC fibers are approximately ten times smaller than the silica fibers. There is some indication a slight $\xi$ dependency on error exists which will be investigated in future work. Regardless, these results demonstrate the accuracy of this novel implementation of TET using a reduced sample size. In this case, the results for $Si0_2$ and $SiC$ demonstrate that for samples with sufficiently low $\xi$, the $\Gamma_2$ scheme produces more accurate values for the same data set with a lower uncertainty. However, great care must be taken when testing samples to ensure $\xi$ is not so large such that the $\alpha$ and $k$ parameter coupling affects the fitted results. It still may be the case that the $\Gamma_1$ scheme is best used to avoid potential parameter coupling if the thermal radiation effect magnitude is unknown.

\begin{figure}[H]
\centering

\begin{subfigure}[t]{0.50\textwidth}
  \centering
  \includegraphics[width=\linewidth]{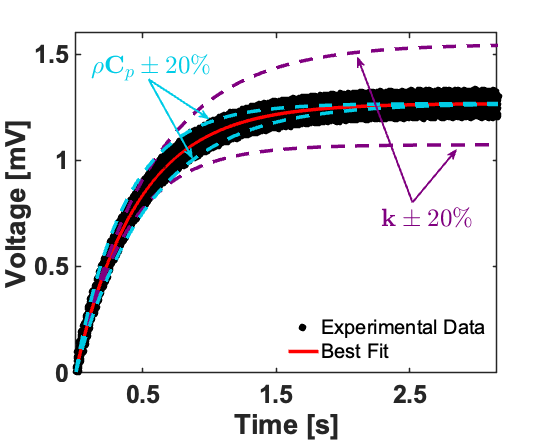}
  \subcaption{}\label{fig:pCp_pm}
\end{subfigure}\hfill
\begin{subfigure}[t]{0.50\textwidth}
  \centering
  \includegraphics[width=\linewidth]{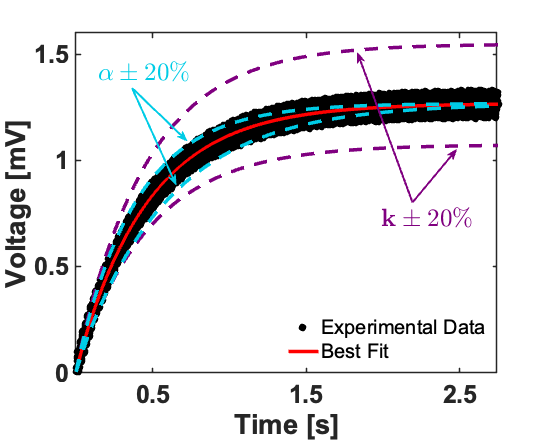}
  \subcaption{}\label{fig:alpha_pm}
\end{subfigure}

\caption{Best fit plots of a selected SiC sample that shows the variation in the temperature solution based on a 20\% increase/decrease in the fitted thermal property for the  (a) $\Gamma_1$ fitting scheme and the  (b) $\Gamma_2$ fitting scheme)}
\label{fig:SiC pm results}
\end{figure}

{Figure~\ref{fig:SiC pm results} shows data for a selected SiC sample with the best fit line corresponding to the determined thermal properties. Both plots include temperature solutions based on a fit property 20\% larger and smaller than the fit result. These curves provide additional insight to parameter sensitivity and the relationship between the two fitting schemes. The differences between Figure~\ref{fig:pCp_pm} and Figure~\ref{fig:alpha_pm} visually confirm that regardless of the diffusivity relationship, $\Gamma_1$ and $\Gamma_2$ are distinct in terms of parameter estimation. Although both schemes demonstrate overlap in the sensitivity of the two fitting parameters, Figure~\ref{fig:alpha_pm} shows a larger dependence on the thermal conductivity at earlier times compared to Figure~\ref{fig:alpha_pm}. This indicates that the thermal conductivity magnitude will have a larger effect on the fit diffusivity which will also present itself as increased parameter coupling. This effect becomes more pronounced at large $\xi$ which results in differences in the parameter uncertainty relationships between the two schemes as detailed in the next section.

\subsection{Uncertainty numerical study}

{As mentioned in previous sections, the relationship between $\xi$ and parameter identifiability has implications on the fitting uncertainty. And, with direct control of the length to alter $\xi$, the experiment can be tuned to obtain a low uncertainty while mitigating other sources of error arising at small lengths. Since $\xi$ and uncertainty are linked, we performed two uncertainty studies using the established formalism.}

\begin{figure}[H]
\centering

\begin{subfigure}[t]{0.48\textwidth}
  \centering
  \includegraphics[width=\linewidth]{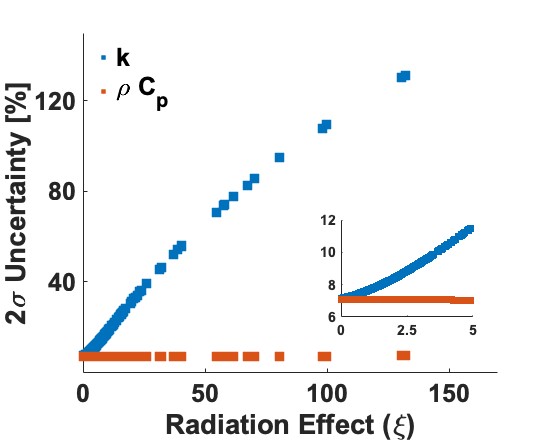}
  \subcaption{}\label{fig:pCp and k}
\end{subfigure}\hfill
\begin{subfigure}[t]{0.48\textwidth}
  \centering
  \includegraphics[width=\linewidth]{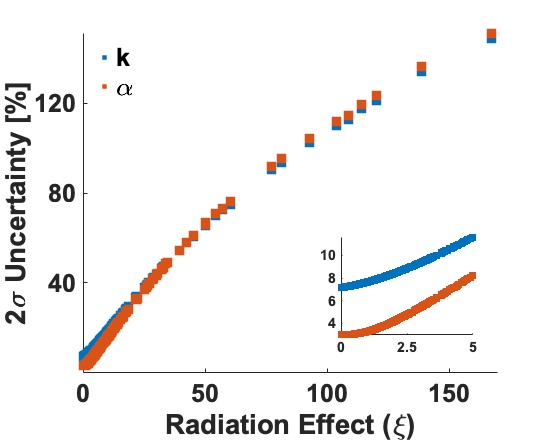}
  \subcaption{}\label{fig:alpha and k}
\end{subfigure}

\caption{Multi-parameter uncertainty numerical study results for (a) $\Gamma_1$ and (b) $\Gamma_2$.}
\label{fig:uncert num study}
\end{figure}

{The first study characterizes how uncertainty varies for the fit properties in both fitting schemes when emissivity is known or negligibly small. The variables governing the transient voltage rise in Equation~\eqref{eq:TET_gov_eqn} were randomized within a representative range of materials tested by TET. The only fixed variables were the temperature coefficient of resistance and the sample initial temperature. Artificial transient data was generated across 1000 trials and noise was added to better represent experimental data. The generated data for each trial was curve fit and plotted against $\xi$. Each trial had their initial guesses randomly selected to ensure any parameter degeneracies were apparent in the data. Note that the control parameter uncertainties are the same as in the formalism, except for diameter. For this study we selected the diameter uncertainty to be 3.34\%, an average of the $SiO_2$ and SiC uncertainties. Figure~\ref{fig:uncert num study} illustrates the differences between the fitting schemes. $\Gamma_1$ demonstrates a significant dependence of uncertainty for conductivity on $\xi$ with negligible effect on volumetric heat capacity. This is consistent with the sensitivity analysis in Figure~\ref{fig:2D sens plots silica} and Figure~\ref{fig:2D sens plots SiC}, in which the trend for $\rho C_p$ remained constant, regardless of length. For the $\Gamma_2$ scheme, we see a different trend in which at low $\xi$, the uncertainties of diffusivity and conductivity are distinct but become equivalent as $\xi$ increases. This is also consistent with the sensitivity analysis, where it was identified that the trends of $k$ and $\alpha$ become indistinguishable at large $\xi$ which results in poor parameter identifiability. These plots indicate that both schemes are viable options with a caveat. The $\Gamma_2$ scheme has a lower uncertainty for small $\xi$, but parameter coupling is likely to impact results for larger $\xi$. Thus, when the $\xi$ is unable to be approximated, the $\Gamma_1$ scheme is the safer option.}

{However, as is often the case in TET, the emissivity is an unknown parameter. Therefore, a second numerical study was performed to analyze the impact $\xi$ has on the uncertainty of the iterative fitting procedure. In this case, the thermal properties and emissivity are generated and applied to two distinct samples. Then, the length and diameter are varied for several combinations of low and high thermal radiation effect. The iterative fitting procedure was then applied to the artificial data and plotted for both fitting schemes.}

\begin{figure}[H]
\centering

\begin{subfigure}[t]{0.48\textwidth}
  \centering
  \includegraphics[width=\linewidth]{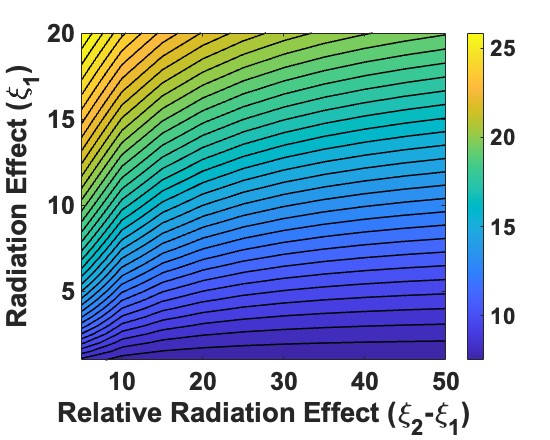}
  \subcaption{}\label{fig:k_iter}
\end{subfigure}\hfill
\begin{subfigure}[t]{0.48\textwidth}
  \centering
  \includegraphics[width=\linewidth]{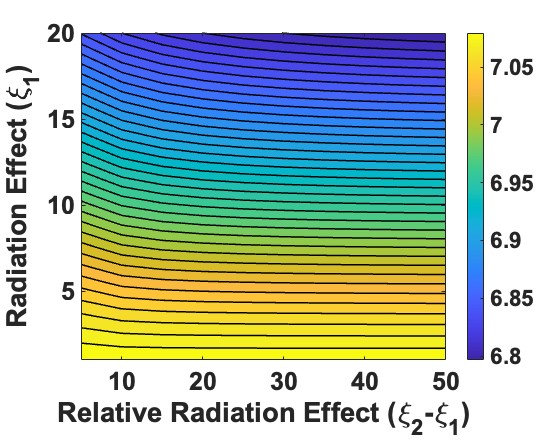}
  \subcaption{}\label{fig:pCp_iter}
\end{subfigure}
\begin{subfigure}[t]{0.48\textwidth}
  \centering
  \includegraphics[width=\linewidth]{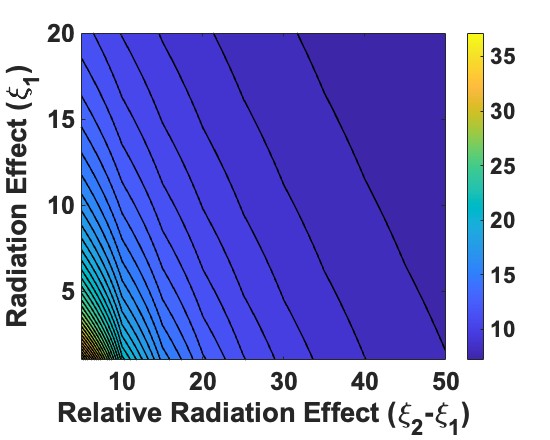}
  \subcaption{}\label{fig:emiss_iter}
\end{subfigure}

\caption{Multi-parameter iterative fitting uncertainty numerical study results for the $\Gamma_1$ scheme for (a) $k$ (b) $\rho C_p$ and (c) $\epsilon$}
\label{fig:iter uncert num study}
\end{figure}

In Figure~\ref{fig:iter uncert num study}, the results of the iterative fitting procedure are plotted so that optimal sample selection can be elucidated. The y-axis represents the thermal radiation effect of the shorter sample responsible for thermal property evaluation and the x-axis is the difference in the thermal radiation between the longer sample ($\xi_2$) and the shorter sample ($\xi_1$). This variable was chosen to demonstrate how the uncertainty changes with respect to proximity in length of the two samples used for iterative fitting. Figure~\ref{fig:k_iter} shows that with increased $\xi_1$, uncertainty of thermal conductivity is far more sensitive to $\xi_2 - \xi_1$. Also, with lower $\xi_1$ values, a smaller $\xi_2 - \xi_1$ is required to obtain a reasonable level of uncertainty. As expected from the sensitivity analysis, Figure~\ref{fig:pCp_iter} shows there is negligible change in uncertainty regardless of $\xi_1$ and $\xi_2 - \xi_1$. For uncertainty variation in Figure~\ref{fig:emiss_iter}, a low uncertainty for emissivity is best achieved at large $\xi_1$ due to the increase in sensitivity to $\epsilon$ with increased $\xi_1$. The results for the $\Gamma_2$ fitting scheme are shown in the supplementary materials document. These studies can be realized as a prescriptive tool for TET fitting when using a multi-parameter approach. By calculating $\xi$ for the material of interest, the proper length can be selected based on approximations of the thermal properties and careful selection of the difference of sample lengths to ensure the fit is well-conditioned. In this way, the inevitable precision errors from the use of small sample lengths, chosen to negate thermal radiation effects, are avoided.

% \section{Conclusions}
\section{Summary and outlook}

In this work, we presented a novel implementation of the TET to perform multi-parameter fitting for two different schemes. Through the use of an iterative curve fitting procedure, we have substantially reduced the required sample size to sufficiently account for environmentally induced errors. Two different materials were used to validate the novel approach and it was found that approximately 10\% error or smaller was obtained for the average of each property. Adaptation of TET also led to the fitting capabilities of an additional parameter, volumetric heat capacity. This creates an alternative to conventionally used methods such as DSC, which can be complex and error prone when determining the specific heat of complex fiber morphologies. Finally, two uncertainty studies were performed which demonstrate the advantages of the fitting schemes and provide a tool to select proper sample lengths to effectively determine the intrinsic thermal properties without errors from thermal radiation.

\label{sec5}
% {The improved methodology now only requires testing two samples as opposed to ten or more as in previous approaches for the TET full model \cite{xing_thermal_2014_3omega,munro_thermal_2016,xing_thermophysical_2014}. This has the potential to substantially reduce the time required to confidently report thermal properties, especially when an annealing step is required for resistance stabilization. Additionally, the fitability of volumetric heat capacity was established. This is the first robust demonstration of fitting for $\rho C_p$ using the TET full model. The ability to determine this property is an especially useful alternative when working with fibers having complex morphologies which make it difficult to determine specific heat through conventional means such as with differential scanning calorimetry (DSC).}

\section{Acknowledgments}
\label{sec6}
This material is based upon work supported under an University Nuclear Leadership Program Graduate Fellowship. The work performed in the University of Pittsburgh Nanofabrication and Characterization Core Facility ($RPID:SCR\_05124$) and the services and instruments used in this project were graciously supported, in part, by the University of Pittsburgh.

\section{Declarations}
The authors declare that they have no known competing financial interests or personal relationships that could have appeared to influence the work reported in this paper. Any opinions, findings, conclusions or recommendations expressed in this publication are
those of the author(s) and do not necessarily reflect the views of the Department of Energy Office of Nuclear Energy. Data used for this work are avaible from the corresponding author upon reasonable request.

\section{Appendix}
\subsection{Uncertainty formalism}

% maybe should include a table like troy used to list all of the uncertainties for the relevant parameters

% Need to determine where the uncertainties come from which are used in the analysis
% For emiss fitting
% pCp assumed 20%
% k assumed 20%
% L-->look at keyence user manual
% I-->look at power source user manual
% Vo--> comes from combining all data points and determine the standard deviation
% D-->look at SEM user manual
% To-->uncertainty would be that of the k-type thermocouple
% TCR-->not sure about this one yet...maybe monte carlo invovled...yeah think i will have to use MC

% I think uncertainty might need to be included later since it has to involve a discussion of the iterative curve-fitting process

% apply uncertainty analysis and uncertainty numerical studies using covariance method
\indent{The uncertainties of the fit parameters were estimated using an error propagation formula. This formula accounts for the uncertainties of all input parameters, the quality of the nonlinear least-squares fitting, and the experimental noise. Although originally designed to determine the uncertainty of TDTR and FDTR measurements \cite{yang_uncertainty_2016}, its application extends to any derivation using a similar fitting scheme.}

\indent{Depending on the fitting scheme, the properties $\rho C_p$, $\alpha$, and $k$ are fit from the $\Delta V(t)$ data through an iterative fitting process which can be mathematically represented as the minimization of the sum of squared differences between experimental data and model predicted data}

\begin{equation}
R = \sum_{i=1}^{M}\left[ v_i(t_i) - f\!\left(t_i, X_U,C\right) \right]^2
\end{equation}

\noindent{where $v_i(t_i)$ is the measured voltage data point corresponding to a time point $t_i$, $C$ is a column vector which represents the control parameters, $X_U$ represents the column vector of unknown parameters to be fit, and $f$ is the corresponding value determined by the thermal model. The vector parameters vary depending on the fitting scheme and the data set used in the iterative fitting process. The process involves two steps, one in which the emissivity is fit, denoted by $X_{{U}_{\epsilon}}$ and $C_{\epsilon}$} and the second in which the thermal properties are fit, denoted by $X_{{U}_{k}}$ and $C_k$. For the shorter sample (small $\xi$ data set), these are $X_{{U}_{k}}=(\rho C_p,k)^T$ or $X_{{U}_{k}}=(\alpha,k)^T$ and $C_{k}=(L,I,V_0,D,\eta,\epsilon,T_0)^T$ depending on the fitting scheme, $\Gamma_1$ or $\Gamma_2$. For the longer sample (large $\xi$), $X_{{U}_{\epsilon}}=(\epsilon)^T$ and $C_{k}=(\rho C_p,k,L,I,V_0,D,\eta,T_0)^T$. The covariance matrix of $X_{U}$ is derived and is expressed by Var[$\hat{X}_{U}$]}

\begin{equation}
\mathrm{Var}[\hat{X}_U] =
\begin{bmatrix}
\sigma_{x_1}^{2} & \mathrm{cov}[x_1,x_2] & \mathrm{cov}[x_1,x_3] & \cdots \\
\mathrm{cov}[x_2,x_1] & \sigma_{x_2}^{2} & \mathrm{cov}[x_2,x_3] & \cdots \\
\mathrm{cov}[x_3,x_1] & \mathrm{cov}[x_3,x_2] & \sigma_{x_3}^{2} & \cdots \\
\vdots & \vdots & \vdots & \ddots
\end{bmatrix}
\end{equation}

\noindent{where the diagonal elements ($\sigma_{x_1}^{2}$,$\sigma_{x_2}^{2}$,...) are the variances of the unknown parameters. The square root of these elements are taken as the $\pm 2\sigma$ (95\% confidence interval) unknown parameter uncertainty. Additional information regarding the uncertainty derivation and specific property uncertainties can be found in the supplementary materials document.

\bibliographystyle{ieeetr}
\bibliography{TET_bib.bib}

@article{yang_uncertainty_2016,
	title = {Uncertainty analysis of thermoreflectance measurements},
	volume = {87},
	issn = {0034-6748, 1089-7623},
	url = {https://pubs.aip.org/rsi/article/87/1/014901/367530/Uncertainty-analysis-of-thermoreflectance},
	doi = {10.1063/1.4939671},
	language = {en},
	number = {1},
	urldate = {2025-01-14},
	journal = {Review of Scientific Instruments},
	author = {Yang, Jia and Ziade, Elbara and Schmidt, Aaron J.},
	month = jan,
	year = {2016},
	pages = {014901},
}

@article{liu_characterization_2014,
	title = {Characterization of {Thermal} {Transport} in {One}-dimensional {Solid} {Materials}},
	issn = {1940-087X},
	url = {https://www.jove.com/t/51144/characterization-of-thermal-transport-in-one--dimensional-solid-materials},
	doi = {10.3791/51144},
	language = {en},
	number = {83},
	urldate = {2024-06-10},
	journal = {Journal of Visualized Experiments},
	author = {Liu, Guoqing and Lin, Huan and Tang, Xiaoduan and Bergler, Kevin and Wang, Xinwei},
	month = jan,
	year = {2014},
	pages = {51144},
}

@article{xing_thermophysical_2014,
	title = {Thermophysical property measurement of electrically nonconductive fibers by the electrothermal technique},
	volume = {25},
	copyright = {http://iopscience.iop.org/info/page/text-and-data-mining},
	issn = {0957-0233, 1361-6501},
	url = {https://iopscience.iop.org/article/10.1088/0957-0233/25/11/115604},
	doi = {10.1088/0957-0233/25/11/115604},
	language = {en},
	number = {11},
	urldate = {2024-06-10},
	journal = {Measurement Science and Technology},
	author = {Xing, Changhu and Munro, Troy and Jensen, Colby and White, Benjamin and Ban, Heng and Copeland, Cameron G and Lewis, Randolph V},
	month = nov,
	year = {2014},
	pages = {115604},
}

@article{guo_thermal_2007,
	title = {Thermal characterization of microscale conductive and nonconductive wires using transient electrothermal technique},
	volume = {101},
	issn = {0021-8979, 1089-7550},
	url = {https://pubs.aip.org/jap/article/101/6/063537/1028206/Thermal-characterization-of-microscale-conductive},
	doi = {10.1063/1.2714679},
	language = {en},
	number = {6},
	urldate = {2024-06-10},
	journal = {Journal of Applied Physics},
	author = {Guo, Jiaqi and Wang, Xinwei and Wang, Tao},
	month = mar,
	year = {2007},
	pages = {063537},
}

@article{xing_thermal_2014,
	title = {Thermal {Characterization} of {Fine} {Fibers} {Using} an {Improved} {Direct} {Electrical} {Heating} {Method}},
	volume = {35},
	issn = {0195-928X, 1572-9567},
	url = {http://link.springer.com/10.1007/s10765-014-1723-0},
	doi = {10.1007/s10765-014-1723-0},
	language = {en},
	number = {8},
	urldate = {2024-06-10},
	journal = {International Journal of Thermophysics},
	author = {Xing, Changhu and Munro, Troy and Jensen, Colby and White, Benjamin and Ban, Heng},
	month = aug,
	year = {2014},
	pages = {1512--1525},
}

@article{munro_thermal_2016,
	title = {Thermal {Property} {Measurement} of {Thin} {Fibers} by {Complementary} {Methods}},
	language = {en},
	author = {Munro, Troy Robert},
	year = {2016},
}

@article{xing_analysis_2013,
	title = {Analysis of the electrothermal technique for thermal property characterization of thin fibers},
	volume = {24},
	copyright = {http://iopscience.iop.org/info/page/text-and-data-mining},
	issn = {0957-0233, 1361-6501},
	url = {https://iopscience.iop.org/article/10.1088/0957-0233/24/10/105603},
	doi = {10.1088/0957-0233/24/10/105603},
	language = {en},
	number = {10},
	urldate = {2024-06-10},
	journal = {Measurement Science and Technology},
	author = {Xing, Changhu and Munro, Troy and Jensen, Colby and Ban, Heng},
	month = oct,
	year = {2013},
	pages = {105603},
}

@article{xing_thermal_2014_3omega,
	title = {Thermal property characterization of fine fibers by the 3-omega technique},
	volume = {71},
	issn = {13594311},
	url = {https://linkinghub.elsevier.com/retrieve/pii/S135943111400502X},
	doi = {10.1016/j.applthermaleng.2014.06.022},
	language = {en},
	number = {1},
	urldate = {2025-02-24},
	journal = {Applied Thermal Engineering},
	author = {Xing, Changhu and Jensen, Colby and Munro, Troy and White, Benjamin and Ban, Heng and Chirtoc, Mihai},
	month = oct,
	year = {2014},
	pages = {589--595},
}

@book{book,
author = {Cole, Kevin and Beck, J.V. and Haji-Sheikh, A. and Litkouhi, Bahman},
year = {2010},
month = {01},
pages = {63-67},
title = {Heat conduction using green's functions, second edition},
journal = {Heat Conduction Using Green's Functions, Second Edition}
}

@article{feng_simultaneous_2009,
	title = {Simultaneous measurements of the specific heat and thermal conductivity of suspended thin samples by transient electrothermal method},
	volume = {80},
	issn = {0034-6748, 1089-7623},
	url = {https://pubs.aip.org/rsi/article/80/6/064901/281966/Simultaneous-measurements-of-the-specific-heat-and},
	doi = {10.1063/1.3153464},
	language = {en},
	number = {6},
	urldate = {2024-06-10},
	journal = {Review of Scientific Instruments},
	author = {Feng, Bo and Ma, Weigang and Li, Zhixin and Zhang, Xing},
	month = jun,
	year = {2009},
	pages = {064901},
}

@article{zhu_novel_2017,
	title = {Novel {Polyethylene} {Fibers} of {Very} {High} {Thermal} {Conductivity} {Enabled} by {Amorphous} {Restructuring}},
	volume = {2},
	copyright = {http://pubs.acs.org/page/policy/authorchoice\_termsofuse.html},
	issn = {2470-1343, 2470-1343},
	url = {https://pubs.acs.org/doi/10.1021/acsomega.7b00563},
	doi = {10.1021/acsomega.7b00563},
	language = {en},
	number = {7},
	urldate = {2024-06-10},
	journal = {ACS Omega},
	author = {Zhu, Bowen and Liu, Jing and Wang, Tianyu and Han, Meng and Valloppilly, Shah and Xu, Shen and Wang, Xinwei},
	month = jul,
	year = {2017},
	pages = {3931--3944},
}

@article{liu_thermal_2015,
	title = {Thermal {Conductivity} of {Ultrahigh} {Molecular} {Weight} {Polyethylene} {Crystal}: {Defect} {Effect} {Uncovered} by 0 {K} {Limit} {Phonon} {Diffusion}},
	volume = {7},
	issn = {1944-8244, 1944-8252},
	shorttitle = {Thermal {Conductivity} of {Ultrahigh} {Molecular} {Weight} {Polyethylene} {Crystal}},
	url = {https://pubs.acs.org/doi/10.1021/acsami.5b08578},
	doi = {10.1021/acsami.5b08578},
	language = {en},
	number = {49},
	urldate = {2024-07-15},
	journal = {ACS Applied Materials \& Interfaces},
	author = {Liu, Jing and Xu, Zaoli and Cheng, Zhe and Xu, Shen and Wang, Xinwei},
	month = dec,
	year = {2015},
	pages = {27279--27288},
}

@article{moon_note_2012,
	title = {Note: {Thermal} conductivity measurement of individual poly(ether ketone)/carbon nanotube fibers using a steady-state dc thermal bridge method},
	volume = {83},
	issn = {0034-6748, 1089-7623},
	shorttitle = {Note},
	url = {https://pubs.aip.org/rsi/article/83/1/016103/389989/Note-Thermal-conductivity-measurement-of},
	doi = {10.1063/1.3676650},
	language = {en},
	number = {1},
	urldate = {2025-12-14},
	journal = {Review of Scientific Instruments},
	author = {Moon, Jaeyun and Weaver, Keith and Feng, Bo and Gi Chae, Han and Kumar, Satish and Baek, Jong-Beom and Peterson, G. P.},
	month = jan,
	year = {2012},
	pages = {016103},
}

@article{guo_transient_2007,
	title = {Transient thermal characterization of micro/submicroscale polyacrylonitrile wires},
	volume = {89},
	copyright = {http://www.springer.com/tdm},
	issn = {0947-8396, 1432-0630},
	url = {http://link.springer.com/10.1007/s00339-007-4201-8},
	doi = {10.1007/s00339-007-4201-8},
	language = {en},
	number = {1},
	urldate = {2024-06-10},
	journal = {Applied Physics A},
	author = {Guo, Jiaqi and Wang, Xinwei and Zhang, Lijun and Wang, Tao},
	month = aug,
	year = {2007},
	pages = {153--156},
}

@article{zhang_measurements_nodate,
	title = {Measurements of {Thermal} {Conductivity} and {Electrical} {Conductivity} of a {Single} {Carbon} {Fiber}},
	language = {en},
	author = {Zhang, X and Fujiwara, S and Fujii, M},
}

@article{xing_thermal_2017,
	title = {Thermal characterization of natural and synthetic spider silks by both the 3{$\omega$} and transient electrothermal methods},
	volume = {119},
	issn = {02641275},
	url = {https://linkinghub.elsevier.com/retrieve/pii/S0264127517300771},
	doi = {10.1016/j.matdes.2017.01.057},
	language = {en},
	urldate = {2024-06-10},
	journal = {Materials \& Design},
	author = {Xing, Changhu and Munro, Troy and Jensen, Colby and Ban, Heng and Copeland, Cameron G. and Lewis, Randolph V.},
	month = apr,
	year = {2017},
	pages = {22--29},
}

@article{volklein_method_1984,
	title = {A {Method} for the {Measurement} of {Thermal} {Conductivity}, {Thermal} {Diffusivity}, and {Other} {Transport} {Coefficients} of {Thin} {Films}},
	volume = {81},
	copyright = {http://doi.wiley.com/10.1002/tdm\_license\_1.1},
	issn = {00318965, 1521396X},
	url = {https://onlinelibrary.wiley.com/doi/10.1002/pssa.2210810222},
	doi = {10.1002/pssa.2210810222},
	language = {en},
	number = {2},
	urldate = {2024-06-10},
	journal = {Physica Status Solidi (a)},
	author = {V{\"o}lklein, F. and Kessler, E.},
	month = feb,
	year = {1984},
	pages = {585--596},
}

@article{lee_ultrahigh_2022,
	title = {Ultrahigh strength, modulus, and conductivity of graphitic fibers by macromolecular coalescence},
	volume = {8},
	issn = {2375-2548},
	url = {https://www.science.org/doi/10.1126/sciadv.abn0939},
	doi = {10.1126/sciadv.abn0939},
	language = {en},
	number = {16},
	urldate = {2025-12-14},
	journal = {Science Advances},
	author = {Lee, Dongju and Kim, Seo Gyun and Hong, Seungki and Madrona, Cristina and Oh, Yuna and Park, Min and Komatsu, Natsumi and Taylor, Lauren W. and Chung, Bongjin and Kim, Jungwon and Hwang, Jun Yeon and Yu, Jaesang and Lee, Dong Su and Jeong, Hyeon Su and You, Nam Ho and Kim, Nam Dong and Kim, Dae-Yoon and Lee, Heon Sang and Lee, Kun-Hong and Kono, Junichiro and Wehmeyer, Geoff and Pasquali, Matteo and Vilatela, Juan J. and Ryu, Seongwoo and Ku, Bon-Cheol},
	month = apr,
	year = {2022},
	pages = {eabn0939},
}

@article{liu_thermally_2014,
	title = {Thermally induced increase in energy transport capacity of silkworm silks},
	volume = {101},
	copyright = {http://onlinelibrary.wiley.com/termsAndConditions\#vor},
	issn = {0006-3525, 1097-0282},
	url = {https://onlinelibrary.wiley.com/doi/10.1002/bip.22496},
	doi = {10.1002/bip.22496},
	language = {en},
	number = {10},
	urldate = {2026-05-18},
	journal = {Biopolymers},
	author = {Liu, Guoqing and Xu, Shen and Cao, Ting-Ting and Lin, Huan and Tang, Xiaoduan and Zhang, Yu-Qing and Wang, Xinwei},
	month = oct,
	year = {2014},
	pages = {1029--1037},
}

@article{xu_energy_2014,
	title = {Energy transport in crystalline {DNA} composites},
	volume = {4},
	issn = {2158-3226},
	url = {https://pubs.aip.org/adv/article/4/1/017131/20526/Energy-transport-in-crystalline-DNA-composites},
	doi = {10.1063/1.4863924},
	language = {en},
	number = {1},
	urldate = {2024-06-10},
	journal = {AIP Advances},
	author = {Xu, Zaoli and Xu, Shen and Tang, Xiaoduan and Wang, Xinwei},
	month = jan,
	year = {2014},
	note = {TLDR: This work reports on the synthesis of crystalline DNA-composited films and microfibers, and details the study of thermal energy transport in them, using the transient electro-thermal technique and photothermal technique.},
	pages = {017131},
}

@article{feser_pump-probe_2014,
	title = {Pump-probe measurements of the thermal conductivity tensor for materials lacking in-plane symmetry},
	volume = {85},
	issn = {0034-6748, 1089-7623},
	url = {https://pubs.aip.org/rsi/article/85/10/104903/361912/Pump-probe-measurements-of-the-thermal},
	doi = {10.1063/1.4897622},
	language = {en},
	number = {10},
	urldate = {2024-10-03},
	journal = {Review of Scientific Instruments},
	author = {Feser, Joseph P. and Liu, Jun and Cahill, David G.},
	month = oct,
	year = {2014},
	pages = {104903},
}

@article{jiang_new_2022,
	title = {A new spatial-domain thermoreflectance method to measure a broad range of anisotropic in-plane thermal conductivity},
	volume = {191},
	issn = {00179310},
	url = {https://linkinghub.elsevier.com/retrieve/pii/S0017931022003313},
	doi = {10.1016/j.ijheatmasstransfer.2022.122849},
	language = {en},
	urldate = {2024-08-12},
	journal = {International Journal of Heat and Mass Transfer},
	author = {Jiang, Puqing and Wang, Dihui and Xiang, Zeyu and Yang, Ronggui and Ban, Heng},
	month = aug,
	year = {2022},
	pages = {122849},
}

@article{tang_anisotropic_2021,
	title = {Anisotropic thermal conductivity tensor measurements using beam-offset frequency domain thermoreflectance ({BO}-{FDTR}) for materials lacking in-plane symmetry},
	volume = {164},
	issn = {00179310},
	url = {https://linkinghub.elsevier.com/retrieve/pii/S0017931020335365},
	doi = {10.1016/j.ijheatmasstransfer.2020.120600},
	language = {en},
	urldate = {2024-07-21},
	journal = {International Journal of Heat and Mass Transfer},
	author = {Tang, Lei and Dames, Chris},
	month = jan,
	year = {2021},
	pages = {120600},
}

@article{rodin_simultaneous_2017,
	title = {Simultaneous measurement of in-plane and through-plane thermal conductivity using beam-offset frequency domain thermoreflectance},
	volume = {88},
	issn = {0034-6748, 1089-7623},
	url = {https://pubs.aip.org/rsi/article/88/1/014902/367479/Simultaneous-measurement-of-in-plane-and-through},
	doi = {10.1063/1.4973297},
	language = {en},
	number = {1},
	urldate = {2024-07-21},
	journal = {Review of Scientific Instruments},
	author = {Rodin, David and Yee, Shannon K.},
	month = jan,
	year = {2017},
	pages = {014902},
}

@article{schmidt_frequency-domain_2009,
	title = {A frequency-domain thermoreflectance method for the characterization of thermal properties},
	volume = {80},
	issn = {0034-6748, 1089-7623},
	url = {https://pubs.aip.org/rsi/article/80/9/094901/354767/A-frequency-domain-thermoreflectance-method-for},
	doi = {10.1063/1.3212673},
	language = {en},
	number = {9},
	urldate = {2024-06-28},
	journal = {Review of Scientific Instruments},
	author = {Schmidt, Aaron J. and Cheaito, Ramez and Chiesa, Matteo},
	month = sep,
	year = {2009},
	pages = {094901},
}

@article{wang_three-dimensional_2025,
	title = {Three-dimensional ({3D}) tensor-based methodology for characterizing {3D} anisotropic thermal conductivity tensor},
	volume = {242},
	issn = {00179310},
	url = {https://linkinghub.elsevier.com/retrieve/pii/S0017931025002273},
	doi = {10.1016/j.ijheatmasstransfer.2025.126886},
	language = {en},
	urldate = {2025-04-28},
	journal = {International Journal of Heat and Mass Transfer},
	author = {Wang, Dihui and Ban, Heng and Jiang, Puqing},
	month = jun,
	year = {2025},
	pages = {126886},
}

@article{sheikh_comparison_2024,
	title = {A comparison between {Hankel} and {Fourier} methods for photothermal radiometry analysis},
	volume = {53},
	issn = {2688-4534, 2688-4542},
	url = {https://onlinelibrary.wiley.com/doi/10.1002/htj.23134},
	doi = {10.1002/htj.23134},
	language = {en},
	number = {8},
	urldate = {2025-05-17},
	journal = {Heat Transfer},
	author = {Sheikh, Raza and Pompidou, Quentin and Villarreal, Ezekiel and Horny, Nicolas and Ban, Heng},
	month = dec,
	year = {2024},
	pages = {4268--4284},
}

@article{hamaoui_spatially_2020,
	title = {Spatially localized measurement of isotropic and anisotropic thermophysical properties by photothermal radiometry},
	volume = {128},
	issn = {0021-8979, 1089-7550},
	url = {https://pubs.aip.org/jap/article/128/17/175104/1063042/Spatially-localized-measurement-of-isotropic-and},
	doi = {10.1063/5.0020411},
	language = {en},
	number = {17},
	urldate = {2025-10-03},
	journal = {Journal of Applied Physics},
	author = {Hamaoui, Georges and Villarreal, Ezekiel and Ban, Heng and Chirtoc, Mihai and Horny, Nicolas},
	month = nov,
	year = {2020},
	pages = {175104},
}

@article{corona_recent_2023,
	title = {Recent {Progress} in {Modulated} {Photothermal} {Radiometry}},
	volume = {23},
	issn = {1424-8220},
	url = {https://www.mdpi.com/1424-8220/23/10/4935},
	doi = {10.3390/s23104935},
	language = {en},
	number = {10},
	urldate = {2025-11-13},
	journal = {Sensors},
	author = {Corona, Javier and Kandadai, Nirmala},
	month = may,
	year = {2023},
	pages = {4935},
}

@article{sheikh_photothermal_2025,
	title = {Photothermal radiometry dual measurement technique to reduce uncertainty and improve convergence},
	volume = {169},
	issn = {07351933},
	url = {https://linkinghub.elsevier.com/retrieve/pii/S0735193325013089},
	doi = {10.1016/j.icheatmasstransfer.2025.109882},
	language = {en},
	urldate = {2026-05-18},
	journal = {International Communications in Heat and Mass Transfer},
	author = {Sheikh, Raza and Pompidou, Quentin and Wang, Dihui and Horny, Nicolas and Ban, Heng},
	month = dec,
	year = {2025},
	pages = {109882},
}

@article{wang_spatially_2024,
	title = {Spatially resolved lock-in micro-thermography ({SR}-{LIT}): {A} tensor analysis-enhanced method for anisotropic thermal characterization},
	volume = {11},
	issn = {1931-9401},
	shorttitle = {Spatially resolved lock-in micro-thermography ({SR}-{LIT})},
	url = {https://pubs.aip.org/apr/article/11/2/021407/3282197/Spatially-resolved-lock-in-micro-thermography-SR},
	doi = {10.1063/5.0191073},
	language = {en},
	number = {2},
	urldate = {2024-06-28},
	journal = {Applied Physics Reviews},
	author = {Wang, Dihui and Ban, Heng and Jiang, Puqing},
	month = jun,
	year = {2024},
	pages = {021407},
}

@misc{coi_ceramics_hi-nicalon_type_s,
	title = {{HI-NICALON\texttrademark} {TYPE} {S} Ceramic Fiber},
	author = {{COI Ceramics, Inc.}},
	howpublished = {Manufacturer datasheet},
	url = {https://www.coiceramics.com/assets/docs/fiber/nicalon/Hi-NICALON-Type-S.pdf},
	urldate = {2026-05-28},
	note = {Accessed 28 May 2026},
}

\end{document}